\documentclass[aps, pra, reprint, superscriptaddress, footinbib, floatfix]{revtex4-2}

\usepackage[colorlinks=true,urlcolor=blue]{hyperref}

\usepackage{amsmath,amssymb,bm}
\usepackage{braket}
\usepackage{graphicx}
\usepackage{comment}
\usepackage{tikz}
\usetikzlibrary{decorations.pathmorphing}
\tikzset{snake it/.style={decorate, decoration=snake}}

\usepackage[colorlinks=true]{hyperref}

\graphicspath{{figures/}}

\usepackage{xstring}
\newcommand*{\aref}[1]{%
	\IfBeginWith{#1}{eq:}{Eq.~\eqref{#1}}{}%
	\IfBeginWith{#1}{fig:}{Fig.~\ref{#1}}{}%
	\IfBeginWith{#1}{tab:}{Table~\ref{#1}}{}%
	\IfBeginWith{#1}{appendix:}{Appendix~\ref{#1}}{}%
	\IfBeginWith{#1}{sec:}{Section~\ref{#1}}{}%
}

\usepackage{xcolor}

\begin{document}

\title{Long-wavelength optical lattices from optical beatnotes: theory and applications}
\date{\today}

\author{T. Petrucciani}
\affiliation{Istituto Nazionale di Ottica, Consiglio Nazionale delle Ricerche (CNR-INO), Largo Enrico Fermi 6, 50125 Firenze, Italy} 

\author{A. Santoni}
\affiliation{University of Naples “Federico II”, Via Cinthia 21, 80126 Napoli, Italy} 
\affiliation{European Laboratory for Nonlinear Spectroscopy (LENS), Via N. Carrara 1, 50019 Sesto Fiorentino, Italy}

\author{C. Mazzinghi} 
\affiliation{Istituto Nazionale di Ottica, Consiglio Nazionale delle Ricerche (CNR-INO), Largo Enrico Fermi 6, 50125 Firenze, Italy} 
\affiliation{European Laboratory for Nonlinear Spectroscopy (LENS), Via N. Carrara 1, 50019 Sesto Fiorentino, Italy}

\author{D. Trypogeorgos}
\affiliation{Institute of Nanotechnology, Consiglio Nazionale delle Ricerche (CNR-Nanotec), via Monteroni 165, 73100, Lecce, Italy}

\author{F. Minardi}
\affiliation{Dipartimento di Fisica e Astronomia, Università di Bologna, Viale C. Berti-Pichat 6/2, 40127 Bologna, Italy}
\affiliation{European Laboratory for Nonlinear Spectroscopy (LENS), Via N. Carrara 1, 50019 Sesto Fiorentino, Italy}

\author{M. Fattori}
\email{fattori@lens.unifi.it}
\affiliation{Istituto Nazionale di Ottica, Consiglio Nazionale delle Ricerche (CNR-INO), Largo Enrico Fermi 6, 50125 Firenze, Italy} 
\affiliation{European Laboratory for Nonlinear Spectroscopy (LENS), Via N. Carrara 1, 50019 Sesto Fiorentino, Italy}
\affiliation{University of Florence, Physics Department, Via Sansone 1, 50019 Sesto Fiorentino, Italy}

\author{M. Modugno}
\affiliation{Department of Physics, University of the Basque Country UPV/EHU, 48080 Bilbao, Spain}
\affiliation{IKERBASQUE, Basque Foundation for Science, 48009 Bilbao, Spain}
\affiliation{EHU Quantum Center, University of the Basque Country UPV/EHU, 48940 Leioa, Biscay, Spain}

\begin{abstract}

    We present a theoretical analysis of Beat-Note Superlattices (BNSLs), a recently demonstrated technique for generating periodic trapping potentials for 
    ultracold atomic clouds, with arbitrarily large lattice spacings while maintaining interferometric stability. By combining two optical lattices with slightly different wavelengths, a beatnote intensity pattern is formed, generating, for low depths, an effective lattice potential with a periodicity equal to the wavelength associated to the difference between the wavevectors of the two lattices. We study the range of lattice depths and wavelengths under which this approximation is valid and investigate its robustness against perturbations. We present a few examples where the use of BNSLs could offer significant advantages in comparison to well established techniques for the manipulation of ultracold atomic gases. Our results highlight the potential of BNSLs for quantum simulation, atom interferometry, and other applications in quantum technologies.

\end{abstract}


\maketitle




\section{Introduction}

The coherent evolution of matter waves is at the core of various quantum technologies based on ultracold gases, including atomic clocks \cite{RevModPhys.87.637}, quantum simulators \cite{Bloch2012Quantum}, atom interferometers \cite{RevModPhys.81.1051}, and quantum computers \cite{Saffman2010Quantum}. However, these devices require the manipulation of atomic clouds using external potentials, which inevitably introduce decoherence.

Optical lattices created by retroreflected laser beams have proven to be a powerful tool for the precise control of the external wavefunction of quantum particles \cite{Morsch_2006, Kasevich_1991, Bloch2005Ultracold}. This advantage arises from the fact that their spatial periodicity is determined by the laser wavelength, which can be stabilized with high precision, using optical cavities \cite{Drever1983Laser}. 
Additionally, laser power instabilities induce common-mode fluctuations of the depth of the lattice sites, with minor influence on the coherent evolution of the atoms.
However, the separation between the potential minima in such lattices is constrained to half the wavelength of the laser used. With only a few notable exceptions \cite{Scheunemann2000}, this typically limits the lattice spacing to a maximum of approximately one micron, due to the scarcity of high-power laser sources at longer wavelengths.
When lattice spacings of several microns are required, an alternative approach involves crossing laser beams at small angles \cite{Morsch_2006, Nelson2007556, Albiez_2005, Li:08, Valtolina2020, Hilker}. Additionally, Digital Micromirror Devices (DMDs) \cite{Li-Chung_2015} or more in general Spatial Light Modulators (SLMs) \cite{doi:10.1126/science.aah3778} can be employed to shape trapping potentials arbitrarily. However, these methods lack interferometric stability, as they depend on the quality of the projection optics and are highly sensitive to thermal drifts and mechanical vibrations of optical components.

Recently, a novel technique based on 
Beat-Note Superlattices (BNSLs) has been experimentally demonstrated as a means to create periodic potentials with arbitrarily large spatial periodicity \cite{BNSL}. Using this approach, effective lattice potentials with periodicities of ten and five microns have been realized, enabling precise manipulation of both the spatial and momentum wavefunctions of ultracold atoms \cite{masi21}. 
This method relies on two standard optical lattices with slightly different wavelengths, such that a large but finite number of spatial oscillations is required for the relative positions of their potential minima to rephase. This rephasing distance determines the large periodicity of the overall potential while preserving the interferometric stability of standard optical lattices.
Although the experimental realization of BNSLs has already been demonstrated, a comprehensive theoretical description of the trapping potential generated by this lattice configuration remains to be fully established.

In this work, we address this open question by performing a complete analysis of BNSLs, studying their sensitivity to system parameters such as the amplitude and phase of the individual optical lattices and how these affect their overall stability. In addition, we explicitly demonstrate how BNSLs can be an extremely useful tool in realistic experimental scenarios by considering a number of possible applications. 

The paper is organized as follows. Section II introduces the BNSL potential and its two spatial periodicities arising from the average or the difference between the wavevectors of the two lattices. 
In Section III, we investigate the low-depth regime of the lattices. Through a perturbative analysis, we show that, for the low-lying energy states, the BNSL potential behaves analogously to an optical lattice with a spatial periodicity equal to the beating distance of the two lattices. 
Here, and throughout the paper, the analysis is performed at the single-particle level. By numerically diagonalizing the Schr\"odinger equation, we examine the validity range of this approximation and analyze how deviations in the energy spectrum arise when 
the commensurability condition on the two wavelengths is broken. 
Section IV focuses on the intermediate and high-depth regimes of the lattices. When tunneling between the
cells of the large-spacing effective potential becomes negligible, the relevant energy scales are determined by the band gaps of the overall periodic potential. We provide a comprehensive analysis of how these gaps depend on the relative phase between the two lattices, their amplitudes, and their wavelengths. 
Finally, in Section V, we discuss three key potential applications of BNSLs that may be of interest to the ultracold gases community. 
In the Appendices, we provide the details of the derivations and supplementary discussions.

\section{System}
\label{sec:system}

We consider a particle of mass $m$ subject to a \textit{bichromatic} optical lattice potential
\begin{equation}
    V_{B}(x) = V_1 \sin^2(k_1 x + \phi_1) + V_2 \sin^2(k_2 x + \phi_2) ,
    \label{eq:potential}
\end{equation}
composed by two lattices with amplitudes $V_i$, wave-vectors $k_i = 2\pi/\lambda_i$, and phases $\phi_i$ $(i=1,2)$. 
Here, we choose the wavelengths $\lambda_i$ to fulfill the condition $(n+1)\lambda_1 = n\lambda_2\equiv \lambda$, with $n$ an integer and $\lambda/2$ representing the actual periodicity of the whole potential (see below). 
The case of $n=1$ has been already studied in several experiments \cite{PhysRevA.73.033605}, while in this work we will focus our analysis to the case of $n \gg 1$, such that the spatial periodicity of $V(x)$ is much larger than that of the two lattices. 
This case has been experimentally implemented for the first time in Ref. \cite{BNSL}.

\begin{figure}[t]
    \centerline{\includegraphics[width=\columnwidth]{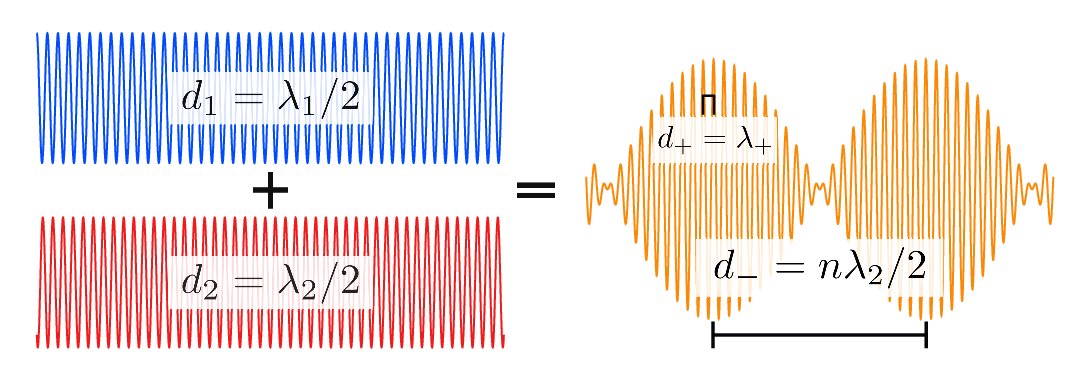}}
    \caption{Sketch of the BNSL potential. Two optical lattices with periodicities $d_i = \lambda_i / 2$ ($i = 1, 2$), satisfying the condition $(n+1)\lambda_1 = n\lambda_2$, are superimposed to produce a fast-oscillating potential with a wavelength of $\lambda_{+} \simeq \lambda_2 / 2$ (here $n=20$). This potential is modulated by a slowly varying periodic amplitude with a period of $d_{-} = n\lambda_2 / 2$. The latter determines the effective super-lattice $V_{\text{eff}}(x)$, as discussed in the  following section [see Eq.~\eqref{eq:veff}].}
    \label{fig:bnsl}
\end{figure}

We focus on the case $V_1 = V_2 \equiv V_{0}$, leaving the general case $V_1 \neq V_2$ to the Appendix \ref{sec:multipot}. By using standard trigonometric transformations the above potential in Eq.~\eqref{eq:potential} can be written as
\begin{equation}
    V_B(x) = V_{0} \left[1- \cos(k_{-} x + \phi_{-}) \cos(k_{+}x + \phi_{+})\right],
    \label{eq:potential_simplyfied}
\end{equation}
where we have defined $k_\pm=k_1\pm k_2$, and $\phi_{\pm}$ similarly. The above expression reveals that, apart from a constant term $V_{0}$, the potential consists of a fast-oscillating term with period $\lambda_{+} = 2\pi/k_{+}$, which is further modulated by a slowly varying periodic amplitude of wavelength $\lambda_{-} = 2\pi/k_{-}=\lambda$. It is also convenient to introduce the Bragg wavevector $k_{B^{+}}\equiv k_+/2$ and the corresponding energy $E_{B^{+}}\equiv \hbar^{2}k_{B^{+}}^{2}/(2m)$, which represents the characteristic momentum and energy scales associated to the fast spatial oscillations. A sketch of the potential is shown in Fig.~\ref{fig:bnsl}.

In the following sections, we will explore various aspects of this potential across different regimes. Initially, we will consider the low lattice depth regime, $V_{0}\ll E_{B^{+}}$, which can be conveniently treated by means of an effective description through a perturbative approach. Then, we will address the intermediate and high depth regimes, where a full numerical approach is required. Hereafter, the example provided corresponds to a configuration with $\lambda_1 = 1013.7$ nm and $n = 20$ \cite{BNSL}, unless stated otherwise.

\section{Perturbative regime}
\label{sec:perturbative}

In this section we focus on the shallow lattice regime, $V_{0}\ll E_{B^{+}}$. Initially, we also set $\phi_{1}=\phi_{2}=0$ in order to simplify the discussion. 
The effect of a phase mismatch between the two lattices will be discussed at the end of this section.

To start with, it is worth noting that in general, when the potential can be factorized into the product of a slowly varying envelope amplitude and a rapidly oscillating periodic component, one can describe the dynamical behavior of the system in terms of an effective Schr\"odinger equation, employing an effective potential. This can be demonstrated under certain hypotheses, as discussed in Ref. \cite{novicenko2019}. Here, we adopt a complementary approach tailored to the specific framework of periodic lattices. This approach is particularly suited for discussing the perturbative regime, employing the standard concepts of Bloch theory for periodic potentials. In particular, it consists in using an envelope function approach along with a perturbative treatment of the fast oscillating component at $k_{+}$, by considering the component at $k_{-}$ as a slow-varying modulation of the amplitude of the former.

\textit{Envelope function approach.}
We start by recalling that the Schrödinger equation governing the evolution of the particle's wave function $\Psi(x,t)$ 
in the presence of a generic periodic potential $V_L(x)=V_L(x+d)$,
\begin{equation}
i\hbar\partial_t\Psi(x,t) = \left[-\frac{\hbar^2}{2m}\nabla^2+V_L(x)\right]\Psi(x,t),
\label{eq:schrod}
\end{equation}
 can be conveniently transformed into an effective equation describing the coarse-grained dynamics on a scale larger than the potential period $d$ (see, for instance, Ref. \cite{morandi2005}). Namely, one can write
\begin{equation}
i\hbar\partial_t\chi_n(x)=\varepsilon_n(-i\nabla)\chi_n(x),
\label{eq:effective}
\end{equation}
where the functions $\chi_n(x)$ are the continuum limit of the amplitudes in the Wannier basis representation, and $\varepsilon_n(k)$ the energy-quasimomentum dispersion relation of the $n-$th Bloch band, with $k\to -i\nabla$ \cite{morandi2005}. Such functions represent the \textit{envelope functions} which provide an effective description of the system, when one is interested in ``macroscopic'' properties on a scale much larger than the lattice spacing. In the following we will focus on the lowest Bloch band, and we will omit the band index for notation easiness. 
 
\begin{figure}[t]
\centerline{\includegraphics[width=0.95\columnwidth]{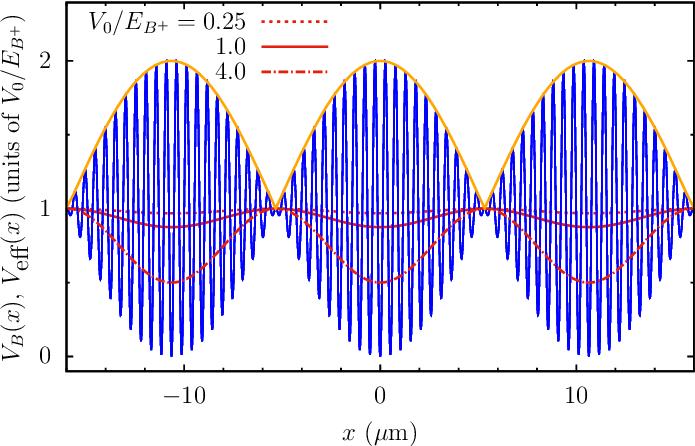}}
\caption{Plot of the BNSL potential $V_B(x)$ (blue solid line) along with the effective potential $V_{\textrm{eff}}$ for three different amplitudes (red lines), $V_{0}/E_{B^{+}}=0.25,1,4$. Notice that energies are rescaled by a (dimensionless) scale factor of $V_0/E_{B^+}$, so that $V_B(x)$ remains invariant in form. The orange line represents the \textit{envelope} of the BNSL, in the terminology of the signal processing analysis (see text).}
\label{fig:fig_bnsl}
\end{figure}

\textit{Perturbative approach.} When the potential is weak, we can use the perturbative approach discussed in Ref. \cite{ashcroft} (see Chapt. 9 therein). Here we focus on the fast oscillating component only, and we rewrite the whole potential as $V(x) = V_{0}\left[1-\alpha\cos(k_{+}x)\right]$. Then, assuming that there is \textit{no near degeneracy} between energy bands, which is the case when $k$ is not too close to the band edges, the dispersion relation can be approximated as ($k\ll k_{+}$)
\begin{equation}
\varepsilon(k) \simeq \frac{\hbar^{2}k^{2}}{2m} + V_{0}\left(1- \frac{V_{0}}{8E_{B^{+}}}\alpha^{2}\right).
\end{equation}
Finally, going back to position space, $k\to -i\hbar\nabla$ [see Eq.~\eqref{eq:effective}] and replacing $\alpha$ with the slowly varying amplitude at $k_{-}$, one can approximate the system's Hamiltonian as $H\equiv\varepsilon(-i\hbar\nabla)=({\hbar^2}/{2m})\nabla^{2} + V_{\textrm{eff}}(x)$ with 
\begin{equation}
V_{\textrm{eff}}(x) = V_{0}\left[1 - \frac{V_{0}}{8E_{B^{+}}}\cos^{2}(k_{-}x)\right]
\label{eq:veff}
\end{equation}
being the \textit{effective potential}. 
Both the BNSL potential and the effective potential $V_{\textrm{eff}}(x)$ are shown in Fig.~\ref{fig:fig_bnsl}, for three different amplitudes $V_0$.
It should be emphasized that, as mentioned before, this result could be also obtained with the method discussed in Ref. \cite{novicenko2019} [see Eq. (21) therein].
Both the BNSL potential and the effective potential $V_{\textrm{eff}}(x)$ are shown in Fig.~\ref{fig:fig_bnsl}, for three different amplitudes $V_0$.

It is also interesting to note that the effective potential $V_{\textrm{eff}}(x)$ can be related to the envelope of the BNSL potential, in the context of signal processing analysis, as discussed in Appendix \ref{sec:analytic_signal}. Specifically, it can be shown that the envelope  of the potential $V(x)\equiv V_{B}(x)/V_{0} -1$ is $|V_{a}(x)|=|\cos(k_{-}x)|$, which corresponds to the orange solid line in Fig.~\ref{fig:fig_bnsl}, modulo a vertical rigid shift. This result can also be inferred directly from the expression for  $V_{B}(x)$ in Eq.~\eqref{eq:potential_simplyfied}. 
Accordingly, the effective potential can be expressed as $V_{B}(x)/V_{0} = 1 - (V_{0}/8E_{B^{+}})|V_{a}(x)|^{2}$.

\begin{figure}[t]
\centerline{\includegraphics[width=0.95\columnwidth]{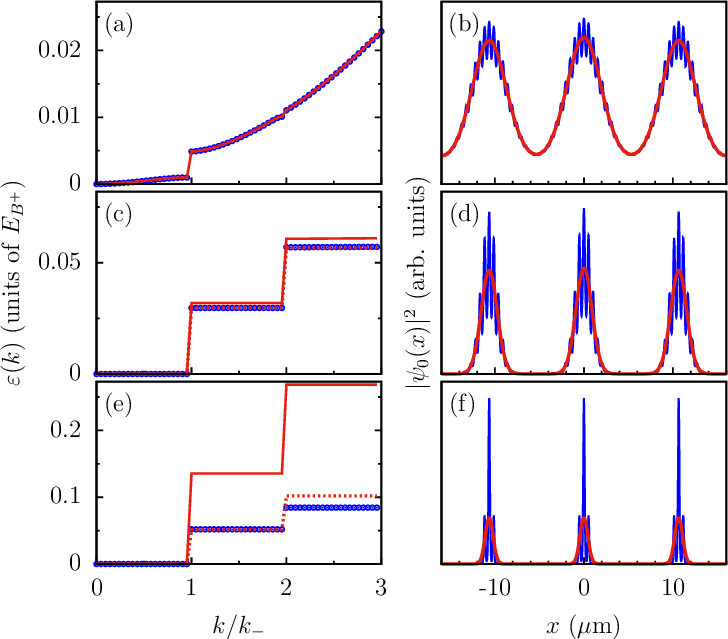}}
\caption{Comparison of the energy dispersion relation (left) and the ground-state density distribution (right) between a BNSL (blue circles) and $V_{\textrm{eff}}$ (red line). 
(a,b) $V_0/E_{B^{+}}=0.25$. (c,d) $V_0/E_{B^{+}}=1$.
At this depth, the analogy starts to fail, but one can recover it with a suitable rescaling of the amplitude of the effective lattice (red dashed line). 
(e,f) $V_0/E_{B^{+}}=4$. 
The structure of the spectrum of a BNSL significantly differs from that of a single-wavelength lattice with the same periodicity, making rescaling no longer applicable.}
\label{fig3}
\end{figure} 

The validity of this approximation is investigated through the numerical analysis shown in Fig.~\ref{fig3}. Here, the energy spectrum and the ground-state wave function of a particle in a BNSL are compared with those in the corresponding effective potential $V_{\textrm{eff}}(x)$, for increasing values of the potential amplitude.
For $V_0 / E_{B^{+}} = 0.25$, the approximation is clearly valid. When $V_0 \simeq E_{B^{+}}$, although the analogy does not fully hold, it is possible to roughly match the first three energy bands of the BNSL with those of $V_{\textrm{eff}}$ by applying a suitable normalization factor, as indicated by the dotted line in Fig.~\ref{fig3}b. Once $V_0$ significantly exceeds $E_{B^{+}}$, the analogy breaks down, as shown in Fig.~\ref{fig3}f for the case $V_0 / E_{B^{+}} = 4$.


\textit{Role of wavelength commensurability and phase mismatch.}
Notably, in the perturbative regime, the properties of the BNSL are insensitive to both the commensurability condition $(n+1)\lambda_1  = n\lambda_2$ and the phases of the two underlying optical lattices. 
Regarding the former, this arises from the generality of the effective potential approach, which extends beyond the specific case of periodic potentials, as discussed in Ref. \cite{novicenko2019}.
As for the phase, it is evident from the perturbative approach leading to Eq.~\eqref{eq:Veff2} that any details regarding the phase $\phi_+$ of the fast spatial modulation are lost, while the relative phase $\phi_-$ only induces a rigid shift in the effective potential.
This result can also be understood as a consequence of the properties of the BNSL envelope, as discussed in Appendix \ref{sec:analytic_signal}. Therefore, we will fix $\phi_-=0$ in the following discussion, without any loss of generality.

To provide a quantitative analysis of these aspects while simultaneously  identifying a figure of merit associated with the validity of the perturbative regime, we introduce the following normalized RMS deviation $\delta\bar{\varepsilon}$ between the spectrum of the effective potential and that of the full BNSL potential,
\begin{equation}
    \delta\bar{\varepsilon}\equiv V_0^{-1}\sum_{k\in{\cal{K}}_3}\sqrt{(\varepsilon_{k}^{BNSL}-\varepsilon_{k}^{eff})^2/N_3},
    \label{eq:rms_deviation}
\end{equation}
which we compute for the first three energy bands, as considered in Fig.~\ref{fig3}, with ${\cal{K}}_3$ representing the set of available $k$-values, and $N_3$ denoting the number of these values (the set is discrete, as it is obtained from a numerical implementation). 
We find that $\delta\bar{\varepsilon} < 10^{-4}$ for $V_0 \leq E_{R^+}$, and it shows no appreciable dependence on the ratio $\lambda_2/\lambda_1$ or on the phase $\phi_+$ of the fast spatial oscillations. This confirms that, within the perturbative regime, the effective potential is indeed a very accurate and highly robust approximation of the BNSL lattice, proving to be insensitive to both phase mismatch and the commensurability of the primary lattices.

\section{General discussion}
\label{sec:general}

\subsection{Intermediate regimes}
\label{sec:intermediate}

One important question to address is whether the BNSL, in the regime of potential depths where it acts like a large-spacing optical lattice, can be sufficiently deep to effectively separate atoms between the lattice sites, thereby minimizing residual tunneling at distances exceeding $d_{-}$. 

Let us introduce the \textit{recoil energy} $E_{R^{-}} = \hbar^2 k_-^2/(2m)$ associated to the effective potential in Eq.~\eqref{eq:veff}, and let us indicate with $s$ the amplitude of the latter, in units of $E_{R^{-}}$. Specifically, $s\equiv V_{0}^2/(8E_{B^{+}})/ E_{R^-}$.
Then, when $V_0\approx E_{B^+}$ and the spacing between sites is large, $n \gg 1$, the effective lattice depth parameter $s\approx(n+1/2)^2/8$ can become very large. As a consequence, the tunneling between neighboring sites of the effective lattice becomes negligible, 
$J_{\textrm{eff}} \simeq J(E_{R^-},s)\equiv(4/\sqrt{\pi})E_{R^-} s^{3/4} e^{-2\sqrt{s}}$ \cite{bloch2008}, 
which is reflected in the flattening of the first energy band, as seen in Fig.~\ref{fig3}. In such a regime, the BNSL can be viewed as an array of potential wells capable of trapping independent clouds of atoms, separated by a distance $d_-$.

\begin{figure}[t]
    \centerline{\includegraphics[width=0.9\columnwidth]{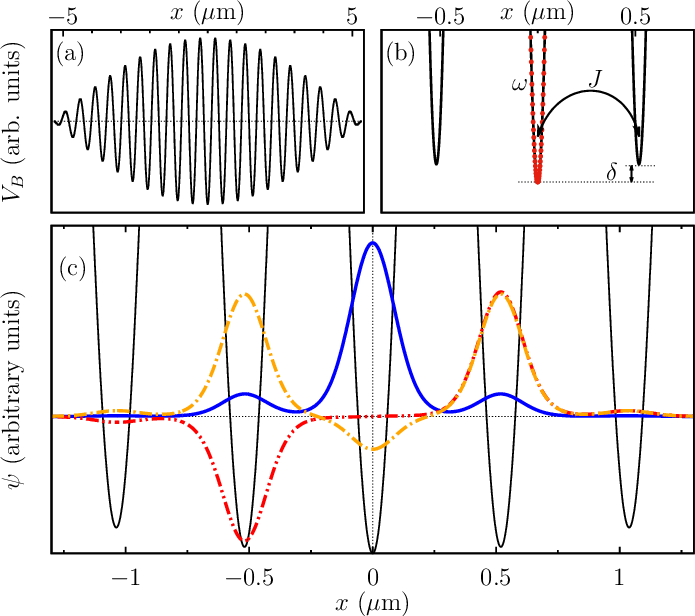}}
    \caption{Plot of the BNSL potential and of its first three eigenstates. 
    (a) The BNSL potential, within a single period of length $d_{-}$. 
    (b) Zoom in on the lowest minimum and its adjacent minima. The red dots indicate the harmonic approximation of each individual well, characterized by the frequency $\omega$. $J$ represents the tunneling between adjacent wells, and $\delta$ denotes the corresponding energy offset (see text). 
    (c) Plot of the ground state (blue) and of the first two excited states (red and orange, respectively) of the BNSL potential (black line; arbitrary units), for $V_0=7E_{B^+}$. Here $\phi_{1}=\phi_{2}=0$.
    } 
    \label{fig4}
\end{figure}
%
Therefore, when $V_0$ exceeds $E_{B^+}$, we can focus the analysis on a single period of the effective potential without any loss of generality, as shown in Fig.~\ref{fig4}a. 
In this scenario, the system is characterized by three energy scales, as indicated in Fig.~\ref{fig4}b: the single-site trapping energy $\hbar\omega$, the tunneling energy $J$ between neighboring sites, and the corresponding energy offset $\delta$.

In order to provide a quantitative estimate of the former two, it is convenient to use as a reference the case of a single, fast-oscillating lattice with wavelength $\lambda_{+}$, see Eq.~\eqref{eq:potential_simplyfied}. Such a potential can be written as $V_+(x)\equiv V_{0}\left[1-\cos(k_{+}x)\right]=s_{+}E_{B^{+}}\sin^2(k_{B^{+}}x)$, with $s_{+}\equiv 2V_0/E_{B^{+}}$. 
In this context, the single-site trapping energy is approximately $\hbar\omega\simeq 2E_{B^{+}}\sqrt{s_{+}}$, which is larger than $E_{B^{+}}$ and increases with $s_{+}$.
The tunneling energy is $J\simeq J(E_{B^+},s_{+})$, which lies below $E_{B^{+}}$ and decreases with $s_{+}$.  
The potential energy difference $\delta$ between neighboring sites, is instead determined by the slowly-varying envelope of the BNSL and it can be approximated as $\delta\simeq V_{B}(2\pi/k_+)-V_{B}(0)$, namely \footnote{This approximation works reasonably well for $n \gg 1$.}
\begin{equation}
\delta \simeq V_0 \frac{2 \pi^2}{(2n+1)^2}.
\label{eq:delta}
\end{equation}
In the intermediate-depth regimes, 
the ratio between $J$ and $\delta$ determines the occupation of each lattice site and the energy levels, given that $\hbar \omega$ is larger than the other energy scales.
In particular, when $\delta \gtrsim J$, only the central site and the two neighboring ones are primarily occupied. This is the case shown in Fig.~\ref{fig4}c, where we plot the wave functions of the ground-state and of the first two excited states of the BNSL potential in Eq.~\eqref{eq:potential_simplyfied}, for $V_0=7E_{B^+}$.

\subsection{Energy gaps}
\label{sec:gap}

\begin{figure}[t]
\centerline{\includegraphics[width=0.9\columnwidth]{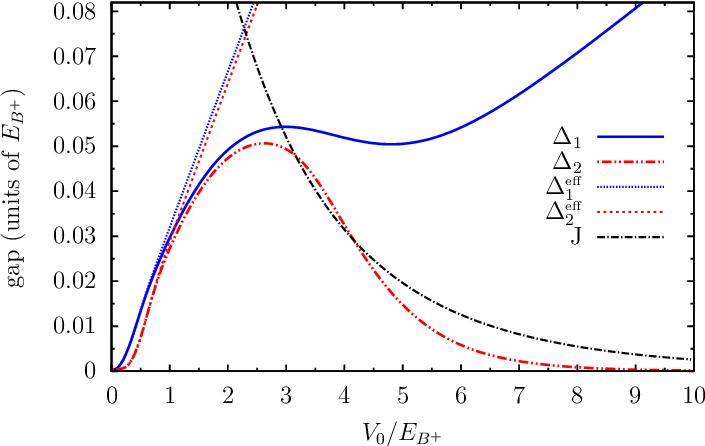}}
\caption{Behavior of the first (blue solid line) and the second (red dashed line) energy gaps of the BNSL, $\Delta_n$ ($n=1,2$), along with those of $V_\textrm{eff}$ (see legend), as a function of $V_0$, for $\phi_1=\phi_2=0$. 
In addition, we plot the tunneling energy $J$ (dotted-dashed line) for a lattice with periodicity $d_+$. Note that the bifurcation of the two energy gaps and their deviation from the $V_\textrm{eff}$ case occur when $J$ becomes comparable to the energy gaps. Here, we use a slightly different expression for $J$ compared to that in Ref. \cite{bloch2008}, namely $J(s_+)=1.43E_{B^+} s_+^{0.98} e^{-2.07\sqrt{s_+}}$ \cite{gerbier2005}, which is more accurate in the regime of low intensities, $s_+<10 E_{B^+}$.}
\label{fig:gap}
\end{figure}

As seen in Fig.~\ref{fig3}e, increasing the lattice intensity flattens the bands, making the spectrum's structure effectively characterized by the arrangement and properties of the band gaps. 
Motivated by this, we now consider the first two energy gaps of the BNSL, which we define as $\Delta_n\equiv\Delta\varepsilon(nk_{-})$ ($n=1,2$), according to Fig.~\ref{fig3}. Their behavior as a function of $V_0$ is shown in Fig.~\ref{fig:gap}, alongside the band gaps $\Delta_n^{\textrm{eff}}$ of the effective potential and the tunneling energy $J$ between neighboring sites of the full BNSL lattice. The former nicely reproduce $\Delta_m$ in the low-depth, perturbative regime, for $V_0\lesssim E_{B^+}$, as expected. The tunneling $J$ provides a typical reference scale. Indeed, it is interesting to note that both gaps $\Delta_n$ of the whole BNSL potential reach a (local) maximum for 
$V_0 \simeq 3E_{B^+}$, where they are of the same order as the tunneling energy $J$. From this point onward, the 
second energy gap decreases towards zero, similarly to $J(s_+)$.

\begin{figure}[t]
\centerline{\includegraphics[width=0.9\columnwidth]{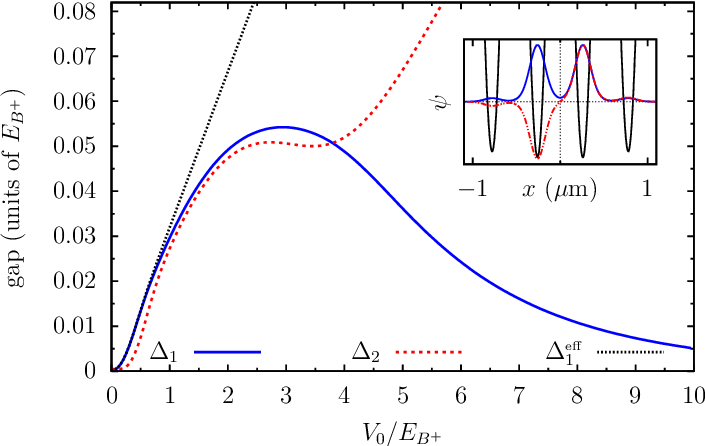}}
\caption{Behavior of the first two band gaps of the BNSL, $\Delta_n$ ($n=1,2$), as a function of $V_0$, for $\phi_1=\phi_2=\pi/2$. Refer to the caption of Fig.~\ref{fig:gap} for a complete description of the symbols. The inset shows the wave function $\psi$ (arbitrary units) for the ground state (blue, solid line) and of the first excited state (red dashed line) of the BNSL potential (black line; arbitrary units), for $V_0=7E_{B^+}$.
}
\label{fig:gap_symmetric}
\end{figure}
It is important to note that the discussion above refers to the case where $\phi_1=\phi_2=0$. Now, let us explore a different scenario, where the fast spatial modulation of the BNSL potential is offset relative to the slowly varying envelope. 
In particular, we will focus on the case 
$\phi_{+}= \pi$, where two degenerate minima are symmetrically displaced with respect to the center of the BNSL envelope, as shown in Fig.~\ref{fig:gap_symmetric}. For sufficiently deep lattices, $V_0=7E_{B^+}$ in the figure, this configuration closely resembles that of a balanced double-well potential, in which the first two eigenstates correspond to nearly degenerate symmetric and antisymmetric wave functions (shown in the inset).
In this case, the behavior of the gaps as a function of the lattice depth $V_0$ is similar to the previous one, but with 
their behaviors interchanged for $V_0\gtrsim 3E_{B^+}$: where $\Delta_1$ was 
increasing and $\Delta_2$ was decreasing, now $\Delta_1$ is decreasing and $\Delta_2$ is increasing.
It is also worth noting that the low-energy behavior does not change, as shown by the fact that $\Delta_1$ nicely matches the first gap of the effective potential. 
This is a general consequence of the robustness of the perturbative approach, which is insensitive to the phase of the lattices and to the commensurability condition, as discussed at the end of Sec. \ref{sec:perturbative}.


\subsection{Effect of quasi-periodicity and phase mismatch}

Building on the previous discussion, we now present a general overview of the roles of commensurability and phase mismatch, focusing on the regime of intermediate lattice amplitudes. This is particularly relevant for assessing whether the control of the BNSL wavelength is critical in the experimental realization of a large-spacing lattice.
In particular, we are interested in how breaking the commensurability condition or introducing a relative phase between the two lattice components of the BNSL affects the distribution of the single-site energies in the large-spacing lattice with period $d_{-}$ (see Fig.~\ref{fig:bnsl}). This might be of interest for high precision measurements applications, like atom interferometry, where the equal depth of the lattice sites is a fundamental condition for the precise operation of the sensor \cite{BNSL, PhysRevA.102.033318}.   

\begin{figure}
\centerline{\includegraphics[width=0.9\columnwidth]{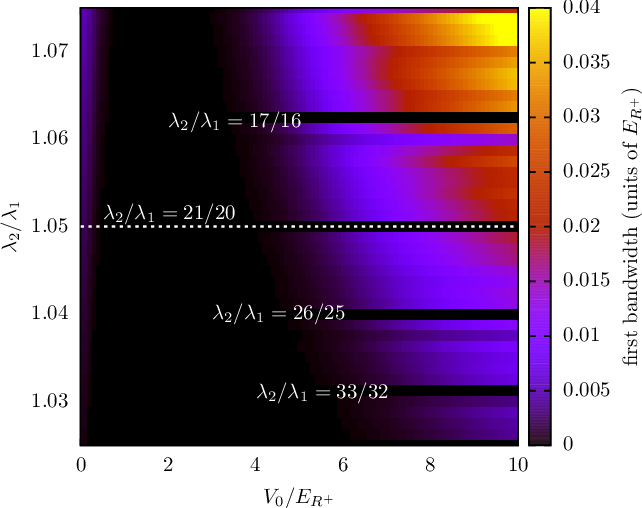}}
\caption{Behavior of the first bandwidth of the BNSL as a function of the lattice amplitude $V_0$ and the wavelength ratio $\lambda_2/\lambda_1$, for $\phi_+=0$. For the ratios of the form $(n+1)/n$, as indicated in the figure ($\lambda_2/\lambda_1=17/16$, $21/20$, $26/25$, $33/32$) the true periodicity of the whole BNSL corresponds to $d_{-}$ as for the effective potential, which explains why the bandwidth does not increase with $V_0$ (see text for a more detailed discussion).}
\label{fig:bw}
\end{figure}
We begin by examining how the width of the first band of the BNSL, $E(k=k_{-})-E(k=0)$, varies as a function of the lattice amplitude $V_0$ and the wavelength ratio $\lambda_2/\lambda_1$, while keeping the phase fixed at $\phi_+=0$, see Fig.~\ref{fig:bw}. The bandwidth provides an estimate of the range over which single-site energies in the large-spacing lattice are distributed.
Overall, the general structure behind the figure can be explained as follows. By increasing $V_0$ and fixed ratio $\lambda_2/\lambda_1$, first the bandwidth decreases due to the presence of the effective slowly-varying potential, which opens a band gap at the Brillouin zone border. This is especially visible in the top left corner of the figure \footnote{Note that the dependence of the bandwidth on the ratio $\lambda_2/\lambda_1$ for low amplitudes, $V_{0}/E_{R^+}\ll1$, does not contradict the validity of the effective potential. It simply reflects the fact that the value of $k_{-}$, which defines the first Brillouin zone, increases with $\lambda_2/\lambda_1$, and consequently, so does the bandwidth (think of the free-particle dispersion relation, in the limit $V_0\to 0$).}.
As $V_0$ is further increased, the system exits the perturbative regime, and the scenario changes depending on the value of $\lambda_2/\lambda_1$. For ``non-commensurate" values, several \textit{mini-gaps} open in the spectrum, including within the first energy band, leading to an overall increase in its width. Notably, this is not the case when the period of the BNSL maintains the same periodicity $d_{-}$ as the effective potential, which occurs when the two wavelengths satisfy the condition $\lambda_2/\lambda_1 = (n+1)/n$. This explains the behavior shown in the figure for $\lambda_2/\lambda_1 = 17/16$, $21/20$, $26/25$, and $33/32$, which do not show any increase in the bandwidth (corresponding to the horizontal black stripes).

\begin{figure}[ht]
\centerline{\includegraphics[width=0.9\columnwidth]{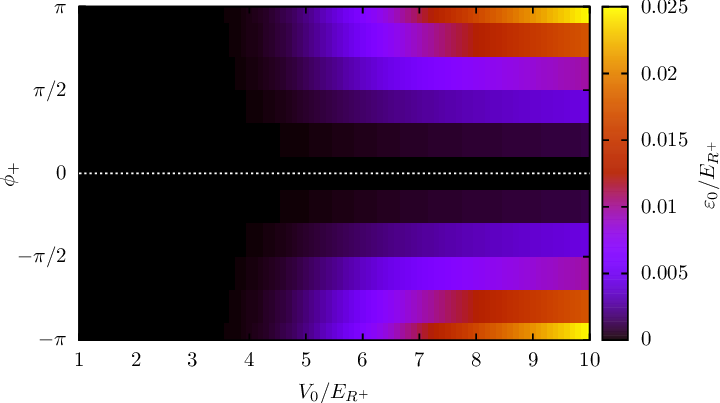}}
\caption{Behavior of the zero-point energy $\varepsilon_{0}$ of the BNSL potential within a single period of length $d_{-}$ (a single well of the large-spacing lattice), as a function of the lattice amplitude $V_0$ and the phase $\phi_+$ of the fast spatial oscillations, for $\lambda_2/\lambda_1=21/20$.
}
\label{fig:sw_phase}
\end{figure}
Additional insights can be obtained by examining how the phase $\phi_+$ of the fast spatial oscillations affects the zero-point energy $\varepsilon_{0}$ of the BNSL potential, within a single period of length $d_{-}$, namely a single well of the large-spacing lattice. This is shown in Fig.~\ref{fig:sw_phase} for the commensurate lattice with $\lambda_2/\lambda_1=21/20$. Although for $V_0 \leq E_{R^+}$ there is no significant variation of $\varepsilon_0$ by varying $\phi_+$, when the depth $V_0 \gg E_{R^+}$ we observe that $\varepsilon_0$ increases for $\phi_+$ approaching $\pi$. This can be explained considering that for $\phi_+ = \pi$ the two lattice sites with minimum energies (see inset of Fig. 6) are shifted to higher energies with respect to the single lattice minimum when $\phi_+= 0$ (see for example Fig. 4). Such shift can be estimated and compared with the results of the numerical simulations in a way similar to what has been done in Eq. (8), calculating $V({\pi}/{k_+}) - V(0) =\delta/4$.

\subsection{Large-depth regime}

We conclude this section by considering the regime of large lattice amplitudes, which is relevant for applications where optical lattices are used to produce strong spatial confinement, such as splitting an atomic sample into separate components -- each trapped in a single lattice well -- and effectively tuning their dimensionality to lower dimensions~\cite{hadzibabic2006}. The following discussion will be useful for the applications of the BNSL presented in the next section.

\begin{figure}[t]
\centerline{\includegraphics[width=0.9\columnwidth]{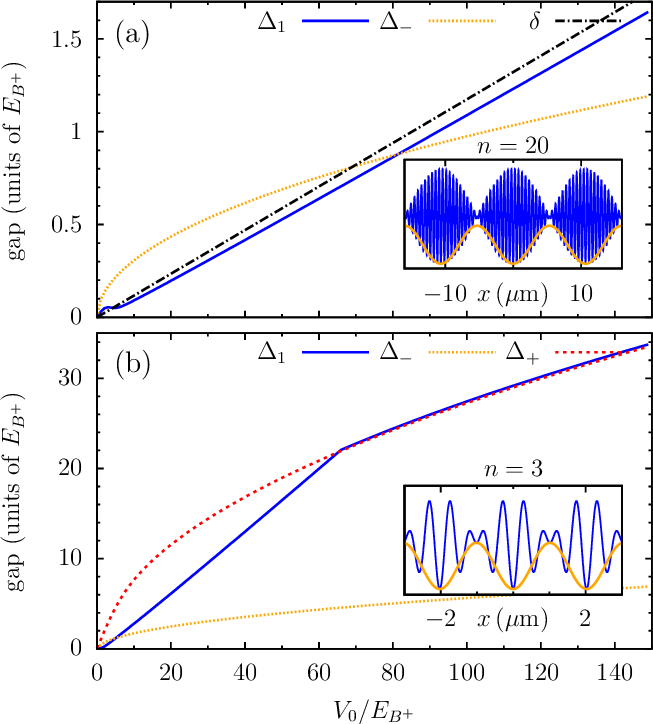}}
\caption{Behavior of the first band gap $\Delta_1$ of a BNSL as a function of $V_0$, for (a) $n=20$ and (b) $n=3$. Also shown are: the gap $\Delta_-$ of a single-wavelength lattice with periodicity $d_-$ and the same ratio of amplitude to recoil energy (orange dotted line); the gap $\Delta_+$ of the corresponding single-wavelength lattice with periodicity $d_+$ [red dashed line in (b)]; 
the quantity $\delta$ in Eq.~\eqref{eq:delta} (dotted-dashed line, see text). The insets show a sketch of the BNSL potential (blue line) and the $V_-(x)$ potential (orange line), in arbitrary units. Here, $\phi_1 = \phi_2 = 0$.}
\label{fig:gap_ld}
\end{figure}

In Fig.~\ref{fig:gap_ld} we show the behavior of the first energy gap $\Delta_1$ of the BNSL, deep into the large-depth regime. For simplicity, we set $\phi_1 = \phi_2 = 0$. 
We start by discussing the case $n=20$ considered so far, shown in panel (a). For $V_0 \gtrsim 1$, the intensity range displayed in the figure corresponds to the tight-binding regime, dominated by the trapping within single wells of the slowly varying envelope (`cells', hereinafter). As shown earlier in Fig.~\ref{fig4}, in this regime the lowest energy state is mostly localized within the deepest well of the rapidly oscillating lattice with wavelength $d_{+}$ (the central `site'), while the first excited states occupy the neighboring sites. As a result, the gap of the BNSL (blue line in Fig.~\ref{fig:gap_ld}a) can be roughly approximated as $\Delta_1 \simeq \delta$ (black dot-dashed line), see Eq.~\eqref{eq:delta}, which corresponds to a linear behavior in $V_0$.

In contrast, the behavior of a single-wavelength lattice is different, with the gap scaling as $\sqrt{V_0}$.
Specifically, in the same figure, we show the gap $\Delta_-$ of a single-wavelength lattice with the same periodicity $d_-$ as the slowly varying envelope (orange dotted line), corresponding to the potential $V_-(x) = V_0 \sin^2(k_- x)$ shown in the inset. It is worth noting that for sufficiently large lattice amplitudes (for $V_0 \gtrsim 80 E_{B^+}$ in the present case), the band gap of the BNSL becomes larger than that of an evenly spaced single lattice.

This feature is especially advantageous for smaller values of $n$, as shown in panel (b) for $n=3$. The figure demonstrates that the BNSL can provide much stronger confinement than a single-wavelength lattice, at equal intensities. Furthermore, for sufficiently large lattice intensity, the gap eventually matches that of a single-wavelength lattice with periodicity $d_+$, given by $\Delta_+ = 2\sqrt{2V_0 E_{B^+}}$ (red dashed line)\footnote{For large $n$, this occurs when $\delta \approx \hbar \omega$, leading to $V_0 \approx 2[({2n+1})/{\pi}]^4 E_{B^+}$.}. In this regime, the ground and first excited states of each cell are fully localized within the central (deepest) site. The practical relevance of these features in experimental applications will be discussed in the next section.

\section{Applications}
\label{sec:applications}

During the last years an increasing number of experimental platforms have required the use of large spacings optical lattices \cite{doi:10.1126/science.aag1635, Gall_Wurz_Samland_Chan_Kohl_2021, PhysRevLett.125.113601}. The main reason for this is the use of light atomic species that have too large tunneling rates in standard half-micrometer spacing optical lattices, or the requirement of single site imaging resolution. In these platforms, BNSLs might represent a valuable alternative to lattices formed with beams crossing at small angles. In addition to this straightforward application, in this section, we explore other situations where BNSLs, could offer improved performances over conventional optical trapping methods.

\subsection{Experiments in lower dimensions}



Optical lattices are routinely used to investigate the physics of ultracold atoms in lower dimensions \cite{hadzibabic2004}, as they provide sufficiently strong confinement in one or two spatial directions. 
Freezing the dynamics along the lattice directions requires two conditions: {\it(i)} the temperature and chemical potential of the atomic sample being smaller than the first energy gap, to avoid populating the excited energy bands; {\it (ii)} the tunneling time between adjacent sites being longer than the time scale of the experiment.

Recently, also ultracold gases of dipolar molecules have been trapped in tightly confined layers formed with optical lattices to suppress inelastic losses and increase their lifetime significantly \cite{ Valtolina2020}. Finally interesting physics emerges when a couple of two dimensional layers are coupled by long range interactions as reported in several studies \cite{Lechner_2013, Cinti_2017, PhysRevA.110.023311}.
%
However, loading a controlled number of lattice sites remains challenging, especially when atoms are produced in low-frequency harmonic potentials or when tight longitudinal confinement and very small lattice spacing are employed. Current techniques mainly rely on site-selective removal or transferring atoms to lattices with progressively shorter periodicities \cite{Sherson_Weitenberg_Endres_Cheneau_Bloch_Kuhr_2010, Ville_2017}.
     
In this section, we discuss how to take advantage of the double spatial periodicity of the BNSL for optimizing the loading of one or two lattice sites, starting from a quantum-degenerate gas in the ground state of a harmonic trap. By increasing the intensity $V_0$ of the BNSL, one can first squeeze the atomic cloud into the minimum of each cell of the effective lattice. Then, by further increasing $V_0$, the atoms can be adiabatically transferred to the lowest energy sites. In other words, in a BNSL, the dynamics along the lattice direction can be effectively frozen either at the scale $d_{-}$ of the slowly varying envelope or at the scale $d_{+}$ of the rapidly oscillating lattice, by progressively increasing the lattice intensity $V_0$. By selecting the appropriate lattice phases, either one or two sites can be conveniently targeted.

\begin{figure}[t]
\centerline{\includegraphics[width=0.99\columnwidth]{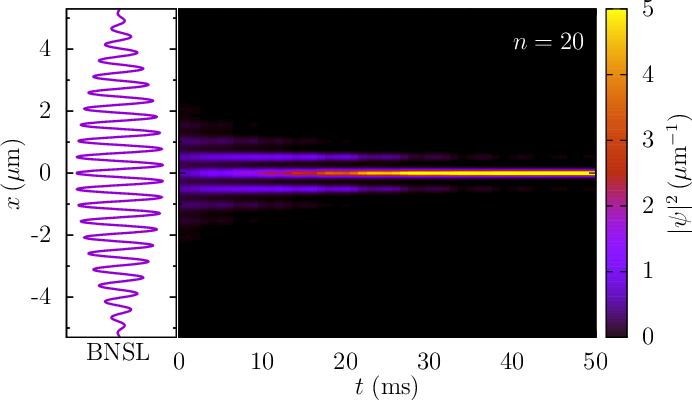}}
\caption{Loading of a BNSL with $n = 20$ from an initial atomic cloud (see text for details). The lattice intensity is ramped linearly from $V_0=1E_{B^+}$ to $V_0=10E_{B^+}$ over a duration of $50$ ms. The figure shows the evolution of the density distribution over time, within a single cell of the BNSL (shown in the inset). Here, $\phi_1 = \phi_2 = 0$.}
\label{fig:loading}
\end{figure}
To provide a specific example, we consider the loading of a non-interacting sample of $^{39}$K into a BNSL potential with $\lambda_1 = 1013.7,\textrm{nm}$ and $n = 20$, as used in the experiment of Ref. \cite{BNSL}.
For simplicity, we assume a one-dimensional setup.
The system is initially prepared in the ground state of a combined potential, consisting of a shallow BNSL with $V_0=1E_{B^+}$ and a harmonic trap with frequency $\omega_x = 2\pi \times 50$~Hz. The harmonic trapping ensures that all atoms remain confined within a single well (a cell) of the effective potential. Then, the BNSL is ramped linearly up to $V_0=10E_{B^+}$ over a time $t_\textrm{ramp} = 50$~ms. As shown in the top panel of Fig.~\ref{fig:loading}, this results in $99\%$ of all the atoms being confined within a single lattice site of the BNSL. Higher loading efficiencies can be achieved using BNSLs with a smaller values of n, that are characterized by larger energy gaps (see Eq. \ref{eq:delta}).

Note that at $V_0 = 10E_{B^+}$, the BNSL already provides the same energy gap that would require higher intensities in a single-wavelength lattice with periodicity $d_{-}$ (see Fig.~\ref{fig:gap_ld}). This naturally leads to a reduction in the radial confinement induced by the finite size of the laser beams,
a reduction that is particularly advantageous in 2D physics experiments where a large number of radial modes below the first energy gap of the lattice are needed.
Similarly, the fact that a BNSL can provide much stronger confinement than a single-wavelength lattice at equal intensities can be especially beneficial for producing an interacting condensate in the 2D regime, for which the condition $\sigma < \xi$ must be fulfilled, where $\xi$ is the healing length and $\sigma$ is the longitudinal size of the condensate \footnote{The longitudinal size $\sigma$ is determined by the effective trapping frequency provided by the lattice potential at each lattice site, which is related to the first energy gap by the relation $\Delta_1 \simeq \hbar \omega_{\text{eff}}$.}.

\subsection{Array of double-well potentials}

Double well potentials have been used to realize atom interferometers using Bose Einstein condensates trapped in two spatial modes \cite{Shin_2004, BerradaNATCOMM2013}. The operation of such devices are strongly affected by trapping instabilities and new configurations immune to noise sources might lead to superior performances.
One interesting solution is offered by a superlattice composed by two lattices with one spatial periodicity equal to half the other \cite{PhysRevA.73.033605}. This configuration realizes an array of double wells where the two modes are spatially separated by half the value of the shorter wavelength, when the optical lattices are realized with beams retro-reflected on a mirror. 
Unfortunately, typical visible or near-infrared radiation leads to sub-micron distances and strong confinement in each well. High densities and consequent strong three body losses limit the maximum atom number that can be manipulated in such potentials.
One way out is to realize large spacing optical lattices using beams crossing at small angles, but this put severe limitations to the stability of the potential that becomes very sensitive to misalignment \cite{EsteveNATURE2008, Trenkwalder, Spagnolli_2017}.

The solution we investigate in this section is the use of a 
pair of large spacing BNSLs. Considering that each BNSL is realized with two lasers one might think that four different wavelengths are required. However we show that the array of double wells can be achieved using only three laser beams whose wavelengths fulfill the 
conditions
\begin{equation}
 \lambda_1= n/(n+1) \lambda_2\,,\quad \lambda_3= n/(n-1) \lambda_2\,,
\end{equation}
with $n$ being an odd integer $\gg 1$.
 The beating between the individual lattices of $\lambda_1$ and $\lambda_2$ creates an effective lattice with lattice spacing $n \lambda_2$, the same occurs for the beating between the lattices of $\lambda_2$ and $\lambda_3$. The two effective lattices with spacing $n\lambda_2$ cancel out when their phases satisfy the balanced double well condition, as we show below.
 In addition, the beating between the lattices of $\lambda_1$ and $\lambda_3$ generates an effective lattice with spacing $n\lambda_2/2$,
 which is exactly half the spacing of the other two.
 
 

\begin{figure}[tb]
    \centerline{\includegraphics[width=\columnwidth]{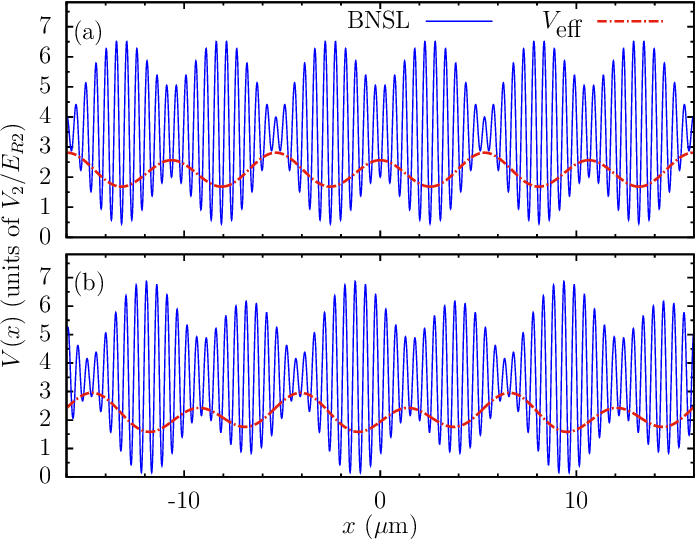}}
    \caption{Plot of the BNSL potential $V_{B3}(x)$ from Eq.~\eqref{eq:superlattice} (blue solid line) alongside the corresponding effective potential $V_{\textrm{eff}}(x)$ from Eq.~\eqref{eq:VeffBL} (red dot-dashed line). (a) Balanced case: $\phi_3 = \phi_2 + \pi/2$, where the effective potential reduces to the simplified expression in Eq.~\eqref{eq:veff_bal}. (b) Unbalanced case: $\phi_3 = \phi_2 + 3\pi/4$. In both cases, the remaining parameters are set to $V_{1}=4V_{2}$, $V_3=0.5V_{1}$, and $\phi_1 = \phi_2 = 0$.}
    \label{DW}
\end{figure}

The main conceptual steps are outlined here, while the details of the calculation are provided in the Appendix \ref{sec:arrays}. We start by considering the potential
\begin{equation} 
\label{eq:superlattice}
V_{B3}(x)= \sum_{i=1,2,3} V_i \sin^2{(k_i x + \phi_i)},
\end{equation}
where $k_i=2\pi / \lambda_i$.
Then, in the perturbative regime, we calculate the corresponding effective potential (see Appendix \ref{sec:arrays}), aiming to express it as a single optical lattice with a slowly varying amplitude. This requirement, which cannot be met for arbitrary $\lambda_i$, is satisfied when the wavelengths fulfill the commensurability condition indicated above.
We first note that lattice 1 and lattice 3 generate a BNSL with a fast spatial modulation with wavevector $k_+^{1,3} = (k_1 + k_3)$, which, for the chosen lattice wavelengths, is exactly equal to $2k_2$. For this reason, it is possible to derive an analytic expression for the sum of the BNSL potential and the second lattice, featuring a fast spatial variation equal to $\lambda_2/2$ and a slowly varying amplitude that defines the effective potential, see Eq.~\eqref{eq:VeffBL}.
In particular, setting $\phi_1 = \phi_2$ and $\phi_3 = \phi_2 + \pi/2$ results in an array of balanced double wells, as shown in Fig.~\ref{DW}a. In this configuration, the effective potential simplifies to
\begin{align}
V_{\textrm{eff}}(x) = & -\frac{(V_1-V_3) V_2 \cos{(k_{-}^{1,3}x)}}{16 E_{R_2}} 
+\frac{V_1 V_3 \cos{(2k_{-}^{1,3}x)}}{16 E_{R_2}},
\label{eq:veff_bal}
\end{align}
where $k_{-}^{1,3}\equiv k_1 - k_3 = 2k_2/n$ and $E_{R_2}=\hbar^2 k_2^2/2m$.

With this lattice configuration in mind, one can first load the atoms into the minima of the effective potential created solely by lattices 1 and 2 (with $V_3 = 0$), which has an amplitude of $V_1 V_2 / 16 E_{R_2}$. Once the atoms are localized, lattice 3 is raised, introducing a barrier at the center of each well with an amplitude of $V_1 V_3 / 16 E_{R_2}$. Simultaneously, the interference between lattices 2 and 3 reduces the amplitude of the effective potential with wavevector $k_{-}^{1,3}$, which vanishes entirely when $V_3 = V_1$. The energy imbalance between the right and left modes within each double well can be tuned by adjusting $\phi_3$ [see Eq.~\eqref{eq:VeffBL}], as shown in Fig.~\ref{DW}b. It is also interesting to note that maintaining the condition $\phi_1 = \phi_2$ while varying the value of $\phi_1$ produces a shift in the fast spatial modulation relative to the slowly varying envelope $V_{\textrm{eff}}(x)$, but it does not alter the overall profile of the effective potential. We expect that this trapping potential configuration might offer superior stability in comparison to former double well trapping potentials and simultaneous operation of several interferometers for residual common noise cancellation.
Finally we would like to stress that array of double wells with large spacings and with interferometric stability might be relevant also for engineering Hubbard couplings for quantum simulation applications as recently reported in \cite{Chalopin_2025}.

\subsection{Kapitza-Dirac interferometry}
The BNSL has already been used to perform multi-mode interferometry on a trapped condensate with time-pulsed lattices, the so-called Kapitza-Dirac (KD) interferometry \cite{smerzi_2014, masi21}. Here the distinct advantage of the BNSL is the possibility to split the original condensate in momentum components finely spaced, $p_j = 2\hbar k_{-} j$ with $j$ integer, 
while suppressing the generation of the ``large'' momenta associated to the individual lattices, i.e. the integer multiples of $2\hbar k_1(k_2)$.
Indeed, while in KD interferometry there is no precise energy resolution, we can exclude the (high) Fourier frequencies required to generate the momentum states at $2\hbar k_1(k_2)$ by elongating the time duration of the pulsed lattice.

To do so, we consider the time-evolution of the single-particle state $\ket{p=0}$, subject to the BNSL for $0<t<T$. 
First, it is useful to recall that $V_B(x) = (V_1/2) [ 1- \cos(2k_1 x) ] + (V_2/2) [ 1- \cos(2 k_2 x +2\phi) ] $ amounts to a sum of translation operators in momentum space, namely:
 \begin{align} 
V_B(x) \ket{p}  &= \frac{V_1 + V_2}2 \ket{p}  - \frac{V_1}{2} \left( \ket{p+2k_1} + \ket{p- 2k_1} \right) \nonumber \\
& - \frac{V_2}{2} \left( e^{i 2\phi} \ket{p+2 k_2} + e^{-i 2\phi} \ket{p-2 k_2} \right),
\end{align}
hereafter we use $\phi=0$ for simplicity.

In interaction representation, the unitary evolution operator $U_I(T, 0)$ is given by the Dyson series that, arrested to the second order, reads:
\begin{widetext}
\begin{align*} 
U_I(T,0) &= 1+ \frac1{i\hbar} \int_0^{T} dt_1 e^{\frac{i}\hbar H_0t_1}V_B(x) e^{-\frac{i}\hbar H_0t_1}  + \frac1{(i\hbar)^2} 
\int_0^T dt_1 \int_0^{t_1} dt_2 e^{\frac{i}\hbar H_0 t_1}V_B(x) e^{-\frac{i}\hbar H_0 (t_1-t_2) } V_B(x) e^{-\frac{i }\hbar H_0 t_2} 
\end{align*}
\end{widetext}
In this case, $H_0$ is the single-particle free Hamiltonian, the eigenstates of $H_0$ are also eigenstates of the momentum, with eigenvalues that can be taken as discrete and numbered. We consider the evolution of the initial state $\ket{0}$ having momentum $\ket{p=0}$. The evolved state is:
\begin{align} 
U_I(T, 0) \ket0  
=&\ket0-  i\sum_m  \ket m \bra m V_B \ket 0 \frac{T}\hbar F\left( \frac{E_m T}\hbar \right) \nonumber \\
&+ i \sum_{m,n} \ket m \bra m V_B \ket n  \frac1{ E_n}  \bra n V_B \ket 0 \frac{T}\hbar \times \nonumber \\
&\left[ F\left( \frac{ E_m T }\hbar \right) - 
F\left( \frac{ (E_m-E_n)T}\hbar \right) \right] 
\end{align}
with $ F(x)  \equiv -i (e^{ix} - 1)/x =  e^{i x/2} \text{sinc}(x/2)$. Clearly the function $F(x)$ enforces energy conservation, since it vanishes for large $x$.

With this result in hand, we consider the transition amplitude towards the large momentum state, e.g. towards  $\ket{2k_1}$, at the leading first order in $V_B$:
\begin{align}
A_{2k_1} \equiv & \bra{2 k_1} U_I(T,0) \ket{0} 
 = 
 -i \frac{V_1 T}{2\hbar}  F\left( \frac{E_{2k_1} T}\hbar \right)
\end{align}
where $E_k \equiv \hbar^2 k^2/(2m)$ is the kinetic energy of a particle with momentum $\hbar k$.

Then we consider the transition amplitude towards small momentum, $\hbar Q = 2\hbar k_1-2\hbar k_2$, at the leading second order in $V_B$:
\begin{align} 
A_Q &\equiv \bra{Q} U_I(T,0) \ket{0} \nonumber\\
&\simeq \, i\frac{T}\hbar  \frac{V_1 V_2}4 
F\left( \frac{ E_Q T }{\hbar} \right)\left(  \frac1{E_{2k_1}} +  \frac1{E_{-2k_2}} \right)  
\label{eq:amplitude_Q}
\end{align}
where we have neglected $F((E_Q-E_{2k_1}) T/ \hbar )$ and $F((E_Q-E_{-2k_2}) T /\hbar )$.

Since the ratio between the two amplitudes is
\begin{align*} 
\frac{A_{2k_1}}{A_Q}  
=&- F\!\left( \frac{E_{2k_1} T}\hbar \right) 
\!\!\left[  \frac{V_2}2
F \!\left( \frac{ E_Q T }{\hbar} \right)\!\!\left(  {2E_{2k_1}} +  \frac{V_2}{2E_{-2k_2}} \right)\right]^{-1}\!\!\!\!, 
\end{align*}
by choosing the time duration of the BNSL such that $ \hbar/E_{2k_1}  \ll T \ll \hbar/E_Q$, we suppress the large-to-small momentum components ratio. With the above condition, $F (E_Q T /\hbar ) \sim 1$ and $F ( E_{2k_1} T / \hbar ) \sim ( E_{2k_1} T /2\hbar)^{-1} $ and
\begin{align}
\left| \frac{A_{2k_1}}{A_Q} \right| \simeq & \frac{\hbar}{E_{2k_1} T} \left(  \frac{V_2}{2E_{2k_1}} +  \frac{V_2}{2E_{-2k_2}} \right).
\end{align}

As detailed in Appendix \ref{sec:perturbation_theory}, the ``small'' momentum component $\hbar Q$ is generated by the four-photons processes whereas the ``large'' momentum component $2\hbar k_1$ by two- photons processes. In principle, one would expect the former to be suppressed with respect to the latter, but, as stated,the opposite holds true due to the Fourier content of the time pulses.

\subsection{Bragg spectroscopy}
Bragg spectroscopy is an important tool to measure the excitation spectrum in many-body quantum gases \cite{Stenger1999}.
In general, two laser beams with unequal frequencies are employed to transfer energy and momentum in a two-photon process. The transferred energy is controlled by adjusting the frequency difference between the two beams; instead, the transferred momentum is usually varied through the angle formed by the two beams. Obviously, the control of the frequency difference is much more precise and accurate than the control of the angle. The BNSL allows to achieve metrological accuracy also for the momentum transfer, since $\hbar k_-$ is determined by the wavelengths of the two BNSL components.
First we notice that, given a pair of counter-propagating beams with different frequencies, we can always find a reference frame where the beams have the same frequency and the lattice they form is at rest. However, in the most general case, for the BNSL the two lattices will be at rest in different reference frames, thus we do not restrict our analysis to lattices at rest.

For Bragg spectroscopy, each time-dependent lattice is obtained by two plane-waves with equal amplitude
 \begin{align*} 
E_j(x,t) \propto e^{-i\omega_j t} \left( e^{i k_j x}  e^{-i\delta_j t} + e^{-i k_j x}  e^{i\delta_j t} \right) \quad j=1,2
\end{align*}
so that, neglecting the dc and uniform terms, the potential reads
\begin{align} 
 V_B(x, t)  =& 
  -\frac{V_1}4 \left[ e^{i 2k_1 x} e^{ - i 2\delta_1 t} + e^{-i 2k_1 x} e^{ i 2\delta_1 t}  \right]  \nonumber \\
  & -\frac{V_2}4 \left[ e^{i 2k_2 x} e^{- i 2\delta_2 t} + e^{-i 2k_2 x} e^{i 2\delta_2 t}  \right] 
\end{align}
In a many-body system, the corresponding Hamiltonian is \cite{stringari_book}:
\begin{align} 
 H_B =& -\frac{V_1}4 e^{- i 2\delta_1 t} \rho_{2k_1}  -\frac{V_2}4 e^{- i 2\delta_2 t} \rho_{2k_2} + \text{h.c.} 
\end{align}
with $\rho_{q} \equiv \sum_{n=1}^N \exp(i q x_n)$.

Again, we consider the transition amplitude from an initial many-body state $\ket 0$ towards a final state with small total momentum, $\hbar Q = 2\hbar k_1-2\hbar k_2$, at the leading second order in $H_B$:

\begin{widetext}
\begin{align} 
A_Q &\equiv \bra{Q} U_I(T,0) \ket{0} \nonumber\\
&\simeq \, i\frac{T}\hbar  \frac{V_1 V_2}4 
F\left( \frac{ (E_Q +2 \hbar \delta_2 - 2 \hbar \delta_1) T }{\hbar} \right)\left[   \frac{ \bra{Q} \rho_{2k_2}^\dagger \ket{n} \bra{n} \rho_{2k_1} \ket 0 }{E_n - 2\hbar\delta_1}  +   \frac{ \bra{Q}  \rho_{2k_1}  \ket{m} \bra{m} \rho_{2k_2}^\dagger  \ket 0 }{E_m +2\hbar\delta_2}   \right]  
\end{align}
\end{widetext}

where we have neglected $F( (E_Q-E_n -2\hbar \delta_2) T / \hbar )$ and $F( (E_Q-E_m -2\hbar \delta_1) T/\hbar)$; $\ket n$ and $\ket m$ are many-body states with total momentum $2\hbar k_1$ and $-2\hbar k_2$, respectively.

With the additional control of the frequency differences $2\delta_1$ and $2\delta_2$, Bragg spectroscopy yields the capability of energy resolution, meaning that we can tune the frequency differences to have $(E_Q +2 \hbar \delta_2 - 2 \hbar \delta_1 )T/\hbar \ll 1$, while $(E_Q-E_n -2\hbar \delta_2) T / \hbar \gg 1$ and $(E_Q-E_m -2\hbar \delta_1) T / \hbar \gg 1$. The same control allows to off-resonantly suppress the excitation of states of large total momentum $\pm 2\hbar k_1$ and $\pm 2\hbar k_2$, originating in two-photon processes, i.e. first-order in $H_B$.

\section{Conclusions and outlook}
\label{sec:conclusion}

We have reported a detailed theoretical analysis of BNSLs over different 
regimes of lattice parameters. 
To begin with, we 
identified the range of depths for which the BNSL behaves like a large spacing effective lattice, providing a rigorous perturbative calculation to 
support this analogy. 
For intermediate lattice depths, we 
studied the dependence of the inter-band energy gaps on the relative phase between the two lattices. 
In addition, we 
quantified the energy mismatch between different effective sites as a function of the ratio of the lattice wavelengths, 
focusing on specific commensurate values. 
For large lattice depths, we 
determined the energy spectrum and studied the transition from the ground state wavefunction, localized in a single effective 
cell of the BNSL, to the single 
site of the rapidly oscillating lattice potential. 

Finally, we have identified three possible scenario where the BNSLs might be relevant for current experiments with ultracold gases. First, 
BNSLs 
could be used as a new tool to load atoms in a single site of an optical lattice, starting from a weakly confined cloud, without any 
particle loss.
In addition, we have shown that a 
pair of BNSLs could be used to create an array of double wells with arbitrarily large spacing and interferometrical stability. Such a configuration 
could find application in atom interferometry with trapped Bose-Einstein condensates and 
might offer a significant improvement 
in the performance of these devices.
Finally, we have derived the range of applications of BNSLs for the realization of Bragg pulses with 
transferred momentum smaller than that associated with a single lattice.

The analysis presented in this work could be a valuable guide for experimentalists 
aiming to implement BNSLs. 
Although the implementation of a beat-note lattice requires the use of two distinct laser sources, its use is certainly advantageous compared to other techniques when the stability of the trapping potential for atoms is a critical requirement. 
Moreover, its realization 
requires less optical access compared to other experimental methods. 
It is therefore realistic to expect that it will be employed in various applications, particularly those related to emerging quantum technologies with ultracold quantum gases.

\begin{acknowledgments}

We acknowledge discussions with M. Landini, E. Kirilov and G. Valtolina. 
We acknowledge financial support by the project SQUEIS of the QuantERA ERA-NET Cofund in Quantum Technologies (Grant Agreement No. 731473 and 101017733) implemented within the European Unions Horizon 2020 Program. We also thank the financial support of the Italian Ministry of Universities and Research under the PRIN2022 project ``Quantum sensing and precision measurements with nonclassical states". MM acknowledges support from Grants No. PID2021-126273NB-I00 funded by MCIN/AEI/10.13039/501100011033 and by ``ERDF A way of making Europe'', and from the Basque Government through Grant No.  IT1470-22. Finally the project has been co-funded by the European Union - Next Generation EU under the PNRR MUR project PE0000023-NQSTI and under the I-PHOQS 'Integrated Infrastructure Initiative in Photonic and Quantum Sciences'.

\end{acknowledgments}

\bibliography{bibliography.bib}

\clearpage
\appendix

\section{Multichromatic potentials}
\label{sec:multipot}

\subsection{Effective potential of a BNSL}

Let us first consider Eq.~\eqref{eq:potential}. By using the bisection trigonometric formula, it can be written as 
\begin{equation}
  V_{B}(x) = \sum_{i=1,2}\frac{V_i}{2}\left[1 - \cos(2k_i x + 2\phi_i) \right],
    \label{eq:potential_2}
\end{equation}
which, by simple algebra, can be conveniently recast as
\begin{align}
    V_B(x) =& \frac{V_1+V_2}{4} \sum_{i=1,2}\left[1 - \cos(2k_i x + 2\phi_i) \right]
    \\
    & -\frac{V_1-V_2}{4} \sum_{i=1,2}(-1)^{i+1}\cos(2k_i x + 2\phi_i).
    \nonumber
    \label{eq:potential_3}
\end{align}
Using the prosthaphaeresis formula on the two pairs of cosines inside each sum, we eventually get
\begin{equation}
    V_B(x) =\frac{V_1+V_2}{2} + A(x) \cos\left(k_{+} x + \phi_{+} - \theta(x)\right),
    \label{eq:potential_4}
\end{equation}
where $k_\pm=k_1\pm k_2$, and similarly $\phi_{\pm}$. 
Here, both $A(x)$ and $\theta(x)$ are slowly varying functions of $x$.
In particular, we have
\begin{equation}
    A(x) = \frac12\sqrt{V_1^2+V_2^2 + 2V_1 V_2\cos(2k_{-} x + 2\phi_{-})},
    \label{eq:amplitude}
\end{equation}
and
\begin{align}
    \cos{\theta(x)} &=  -\frac{V_1+V_2}{2} \frac{\cos{(k_- x +\phi_-)}}{A(x)},
    \label{eq:potential_7}
\\
    \sin{\theta(x)} &= \frac{V_1-V_2}{2} \frac{\sin{(k_- x +\phi_-)}}{A(x)}.
\end{align}
Finally, by means of the perturbative approach of Sec. \ref{sec:perturbative}, we obtain 
\begin{equation}
    V_{\textrm{eff}}(x)= \sum_{i=1}^{2}\left(\frac{V_i}{2}-\frac{V_i^2}{32 E_{B^+}}\right) 
    - \frac{V_1 V_2}{16 E_{B^{+}}}\cos(2k_- x + 2\phi_{-}),
    \label{eq:Veff2}
\end{equation}
which matches Eq.~\eqref{eq:veff} when $V_1=V_2=V_0$.

\subsection{Array of double-well potentials}
\label{sec:arrays}

The above approach can be extended to the case of a potential obtained by superimposing by the three sinusoidal lattices, as in Eq.~\eqref{eq:superlattice}. In particular, we consider the case in which the three wavelengths fulfill the commensurate conditions $\lambda_1 = n/(n+1) \lambda_2$ and $\lambda_3=n/(n-1) \lambda_2$.
Therefore, it is convenient to consider first the BNSL formed by first and the third lattice that, Eq.~\eqref{eq:potential_4}, can be written as
\begin{equation}
    V_B^{1,3}(x) =\frac{V_1+V_3}{2} + A^{1,3}(x) \cos\left(k_+^{1,3} x + \phi_{+}^{1,3} - \theta^{1,3}(x)\right),
\end{equation}
where we have introduced the superscript $1,3$ to identify the lattices the above expression refers to.
Notice that $k_+^{1,3}=k_1+k_3=2 k_2$.
As a consequence, the total potential
\begin{equation}
    V_{B3}(x) =V_B^{1,3}(x) + V_2 \sin^2{(k_2 x +\phi_2)},
    \label{eq:potential_10}
\end{equation}
results from the sum of two potentials characterized by the same fast spatial periodicity of $\lambda_2/2$.
Their sum can be easily calculated analytically. We apply the half angle trigonometric identity on the second term of Eq.~\eqref{eq:potential_10} and then the sum formula of the derived expression to get      
\begin{equation}
    V_{B3}(x) = \frac12\sum_{i=1}^{3} V_{i} + B(x) \cos{(2 k_2 x)} + C(x) \sin{(2 k_2 x)},
    \label{eq:potential_11}
\end{equation}
where 
\begin{equation}
    B(x) = -\frac{V_2}{2} \cos{(2\phi_2)} - A^{1,3}(x) \sin{(\phi_{+}^{1,3} - \theta^{1,3}(x) - \frac{\pi}{2})}
    \label{eq:potential_12}
\end{equation}
and
\begin{equation}
    C(x) = \frac{V_2}{2} \sin{(2\phi_2)} - A^{1,3}(x) \cos{(\phi_{+}^{1,3} - \theta^{1,3}(x) - \frac{\pi}{2})}
    \label{eq:potential_13}
\end{equation}
are slowly varying functions. As a consequence the overall potential can be written, except constant terms, as a single lattice with periodicity $\lambda_2/2$ and amplitude $\sqrt{B^2(x)+C^2(x)}$. According to Eq.~\eqref{eq:veff} the effective potential is $V_{\textrm{eff}}^{1,2,3}(x)=-(B^2(x)+C^2(x))/E_{R_2}$. 
After some algebra, we get
\begin{widetext}
\begin{align}
    V_{\textrm{eff}}(x) = 
    - \frac{1}{16 E_{R_2}}\left[
      V_1 V_2\cos{(k_{-}^{1,3}x + 2\phi_{-}^{1,2})}
    + V_2 V_3\cos{(k_{-}^{1,3}x + 2\phi_{-}^{2,3})}
    + V_1 V_3 \cos{(2k_{-}^{1,3}x + 2\phi_{-}^{1,3})}
    \right],
    \label{eq:VeffBL}
\end{align}
\end{widetext}
modulo a constant term.

\section{Analytic signal analogy}
\label{sec:analytic_signal}

In this section, we draw a useful analogy with analytic signal processing.
An \textit{analytic signal} is a complex-valued function that has no negative frequency components, where the real and imaginary parts are related by an Hilbert transform. Therefore, starting with a real signal -- in our case, an amplitude-modulated lattice potential $V(x)$ -- we can obtain its analytic representation by defining
\begin{equation}
V_{a}(x) \equiv V(x) + i {\cal H}\left[V(x)\right],
\end{equation}
where ${\cal H}\left[\cdot\right]$ represents the Hilbert transform \cite{bracewell,mathworld}. To accurately determine the  envelope of $V(x)$, it is useful to remove the zero frequency component, commonly referred to as the \textit{DC offset}, defined as $V_{DC}=\lim_{L \to \infty} (1/2L) \int_{-L}^{L} V(x) \, dx$. The \textit{envelope} of $V(x)$ is then obtained as $|V_{a}(x)-V_{DC}|+V_{DC}$, which is also known as the \textit{instantaneous amplitude} in case of time-varying signals.

\begin{figure}[ht]
\centerline{\includegraphics[width=0.95\columnwidth]{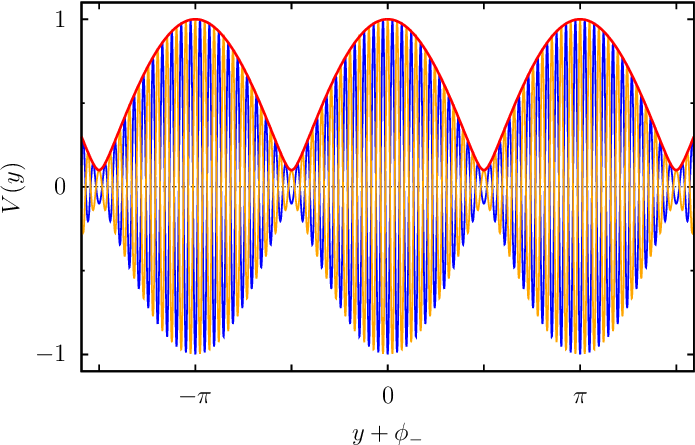}}  
 \caption{Plot of the potential $V(y)$ in Eq.~\eqref{eq:potential_2bis}, for $n=20$, $\alpha=0.1$, and two different values of phase: $\phi_+=0$ (solid blue line) and $\phi_+=\pi$ (solid orange line). The red line depicts the envelope $|V_{a}(y)|$ in Eq.~\eqref{eq:envelope2}.}
 \label{fig:envelope}
\end{figure}

Let us then consider a generic BNSL potential as given in Eq.~\eqref{eq:potential_2}. By defining $V_{1,2}\equiv V_{0}(1\pm\alpha)$, $y\equiv k_{-}x$, we can rewrite it in a dimensionless form,  without loss of generality, as follows:
\begin{align}
  V(y)\equiv& \frac{V_{B}(y)}{V_{0}} = -\frac{1}{2}\left[(1+\alpha)\cos(2ny + \phi_{+}+\phi_{-}) \right.
  \nonumber\\
  &\left. +(1-\alpha)\cos(2(n+1)y +\phi_{+}-\phi_{-}) \right],
    \label{eq:potential_2bis}
\end{align}
where we have disregarded the DC offset $V_{0}$ of the BNSL. By computing the Hilbert transform, the analytic signal representation can be written in compact form, after some algebra, as
\begin{equation}
V_{a}(y+\phi_{-})= -\left[\cos (y)+i \alpha  \sin (y)\right]e^{\displaystyle-2i\theta(y)},
\end{equation}
with $\theta(y)=(n+1)(y+\phi_{-}) +\phi_{+}$, so that the envelope is
\begin{equation}
|V_{a}(y+\phi_{-})|=\sqrt{\cos^{2}(y) + \alpha^{2}\sin^{2}(y)}.
\label{eq:envelope2}
\end{equation}
Notice that the envelope is sensitive only to the amplitude difference between the two optical lattices, proportional to $\alpha$. It does not depend on $\phi_+$, whereas the relative phase $\phi_-$ represents only a rigid shift. An example configuration is shown in Fig.~\ref{fig:envelope}.

At this point, we are able to establish the connection between the above formulation and the effective potential for the perturbative regime.
To begin with , it is straightforward to verify that the above expression \eqref{eq:envelope2} is equivalent to that of the amplitude $A$ in Eq.~\eqref{eq:amplitude}, given by $|V_{a}(y)|=A(y)$. Then, the effective potential in Eq.~\eqref{eq:Veff2} is found to be proportional, modulo a constant DC term, to the square of the amplitude of the corresponding analytic signal:
\begin{equation}
V_{\textrm{eff}}(y)\propto -|V_{a}(y)|^{2}.
\end{equation}

\section{Perturbation theory}
\label{sec:perturbation_theory}

\subsection{Optical potential}
Generally speaking, the optical potential as written in Eq.~\eqref{eq:potential} amounts to the ac Stark shift of the atomic ground state and it arises from second order perturbation theory of the electric dipole Hamiltonian:

\begin{align*}
H_d &  = - \mathbf{d} \cdot \mathbf{\epsilon} \frac12 {\cal E}(\mathbf{R}) \left[ \exp(i \omega t) + \text{c.c.} \right] 
\end{align*}

Indeed, with the evolution operator in interaction representation we find the probability amplitude to remain in the ground state:  
\begin{align*}
\bra{0} U_I(T,0) \ket{0}  =
& 1+   
\frac1{(i\hbar)^2}\sum_n \int _0^T  dt_1 
\int _0^{t_1} dt_2 \\
& 
e^{\frac{i}\hbar E_0 t_1} e^{i \omega  t_1} \frac{\hbar\Omega_n (\mathbf{R})}2  e^{- \frac{i}\hbar E_n t_1} \times \\
& e^{\frac{i}\hbar E_n t_2} e^{-i \omega  t_2} \frac{\hbar\Omega_n (\mathbf{R})}2 e^{- \frac{i}\hbar E_0 t_2}  \\
=&1 -i T \sum_n \frac{\Omega_n^2 (\mathbf{R})}{4\delta_n}  \left[ 
1 - e^{i \delta_n T/2} \text{sinc}(\delta_n T/2)\right]
\end{align*}
where we have defined the ``single-photon'' Rabi frequencies  and detunings
\begin{align*}
\hbar \Omega_n & = \bra{n} \mathbf{d} \ket 0 \cdot \mathbf{\epsilon} {\cal E} (\mathbf{R}), \\
\hbar \delta_n & = \hbar \omega - (E_n - E_0).
\end{align*}

It is convenient to represent the probability amplitude with a diagram:
\begin{center}
\begin{tikzpicture}[thick]
\def\u{1.5}
\node(t1) at (-\u,0) {};
\node (t2) at (-2*\u,0) {};
\node (t3) at (-3*\u,0) {};
\draw (0,0) -- (t1) node [above, midway] {$\ket 0$} ;
\filldraw (t1) circle (0.05) node [below] {$\Omega_n e^{-i \omega t_2}$};
\draw[dashed] (t1) -- (t2) node[midway, above] {$\ket n$};
\filldraw (t2) circle (0.05) node [below] {$\Omega_n e^{i \omega t_1}$};
\draw (t2) -- (t3) node [above, midway] {$\ket 0$} ;
\end{tikzpicture}
\end{center}
It is clear that the first interaction at time $t_2$ absorbs a photon, while the second interaction at time $t_1>t_2$ emits a photon. Notice that time flows from right to left.
For simplicity, we redraw the diagram as:
\begin{center}
\begin{tikzpicture}[thick]
\def\u{1}
\def\p{0.6*\u}
\node(t1) at (-\u,0) {};
\node (t2) at (-2.8*\u,0) {};
\node (t3) at (-3.8*\u,0) {};
\draw (0,0)  -- (t1) node [above, midway] {$\ket 0$} ;
\draw[snake it, <-] (t1)  -- + (\p,-\p);
\filldraw (t1) circle (0.05);
\draw[dashed] (t1) -- (t2) node[midway, above] {$\ket n$};
\draw[snake it, ->] (t2) -- +(-\p,\p);
\filldraw (t2) circle (0.05);
\draw (t2) -- (t3) node [above] {$\ket 0$} ;
\end{tikzpicture}    
\end{center}
The term corresponding to the diagram
\begin{center}
\begin{tikzpicture}[thick]
\def\u{1}
\def\p{0.6*\u}
\node(t1) at (-\u,0) {};
\node (t2) at (-2.8*\u,0) {};
\node (t3) at (-3.8*\u,0) {};

\draw (0,0)  -- (t1) node [above, midway] {$\ket 0$} ;
\draw[snake it, <-] (t2)  -- + (\p,-\p);
\filldraw (t1) circle (0.05);
\draw[dashed] (t1) -- (t2) node[midway, above] {$\ket n$};
\draw[snake it, ->] (t1) -- +(-\p,\p);
\filldraw (t2) circle (0.05);
\draw (t2) -- (t3) node [above] {$\ket 0$} ;
\end{tikzpicture}    
\end{center}
whereby a photon is first emitted and then absorbed, is smaller by a factor $\delta_n/(\omega+ E_n-E_0)$ and usually neglected (``rotating-wave approximation''). 

At sufficiently long time scales, i.e. for $\delta_n T \gg 1$, the sinc function vanishes:
\begin{align*}
\bra{0} U_I(T,0) \ket{0} \simeq& 1 - i T \sum_n \frac{\Omega_n^2 (\mathbf{R})}{4\delta_n} \simeq \exp(-i \Omega_{eff} (\mathbf{R}) T)  \\
\Omega_{eff} (\mathbf{R}) \equiv &  \sum_n \frac{\Omega_n^2 (\mathbf{R})}{4\delta_n}   
\end{align*}

The ``two-photons'' effective Rabi frequency defines the optical potential: $V (\mathbf{R}) \equiv \hbar \Omega_{eff} (\mathbf{R})$.
With multiple fields at different frequencies, the dipole Hamiltonian reads

\begin{align*}
H_d & = - \mathbf{d} \cdot \sum_\alpha  \mathbf{\epsilon}_\alpha \frac12 {\cal E}_{\alpha} (\mathbf{R}) \left[ \exp(i \omega_\alpha t) + \text{c.c.} \right],
\end{align*}

and the above amplitude $\bra 0 U_I(T,0) \ket 0$ contains crossed terms where the photon absorbed is at a frequency different of the photon emitted, corresponding to diagrams
\begin{center}
\begin{tikzpicture}[thick]
\def\u{1}
\def\p{0.6*\u}
\node(t1) at (-\u,0) {};
\node (t2) at (-2*\u,0) {};
\node (t3) at (-3*\u,0) {};

\draw (0,0)  -- (t1);
\draw[snake it, <-] (t1)  -- + (\p,-\p) node [right] {$\omega_\alpha$};
\filldraw (t1) circle (0.05);
\draw[dashed] (t1) -- (t2);
\draw[snake it, ->] (t2) -- +(-\p,\p) node[right] {$\omega_\beta$};
\filldraw (t2) circle (0.05);
\draw (t2) -- (t3);
\end{tikzpicture}    
\end{center}

However, these terms are not energy-conserving and are suppressed by the function $\text{sinc}[(\omega_\alpha - \omega_\beta)T/2]$. Thus, in the stationary case the optical potential is simply obtained summing the potentials generated by the individual frequencies:
\begin{align*}
\Omega_{eff} (\mathbf{R}) \equiv &  \sum_\alpha \sum_n \frac{\Omega_{n,\alpha}^2 (\mathbf{R})}{4\delta_{n,\alpha}} 
\end{align*}

\subsection{Bragg transitions}

Let's first consider a single-frequency standing wave formed by adding two running waves with equal amplitudes ${\cal A}_0$: 
\begin{align*}
\mathbf{{\cal E}}(\mathbf{R}) 
 &=   {\cal A}_0 \left[\exp(i k x) + \text{c.c.}\right]  
\end{align*}.

Alongside the optical potential, arising from considering the two running waves individually, we can calculate the amplitude probability for a Bragg transition, that changes the atomic momentum:

\begin{align*}
\bra{0; p= 2\hbar k } U_I(T,0) \ket{0; p=0}  &= -i T \sum_n \frac{\Omega_n^2}{4\delta_n} \times \\
& \left[
F\left( \frac{\hbar k^2 T}{m} \right)- F\left(\frac{\delta_n T}2\right)\right]  
\end{align*}
with $ \hbar \Omega_n  = \bra{n} \mathbf{d} \ket 0 \cdot \mathbf{\epsilon} {\cal A}_0$. This corresponds to the diagram:
\begin{center}
\begin{tikzpicture}[thick]
\def\u{1}
\def\p{0.5*\u}
\node(t1) at (-\u,0) {};
\node (t2) at (-2*\u,0) {};
\node (t3) at (-3*\u,0) {};

\draw (0,0) node [above] {$0$} -- (t1)  ;
\draw[snake it, <-] (t1)  -- + (\p,-\p) node[right] {$+k$};
\filldraw (t1) circle (0.05);
\draw[dashed] (t1) -- (t2);
\draw[snake it, ->] (t2) -- +(-\p,\p) node[right] {$-k$};
\filldraw (t2) circle (0.05);
\draw (t2) -- (t3) node [above] {$2\hbar k$} ;
\end{tikzpicture}    
\end{center}
showing that it is also a two-photon process
(henceforth we omit the label of the atomic internal state).
.

With the BNSL we have two standing waves at frequencies $\omega_1$ and $\omega_2$. In principle, a two-photon Bragg process could lead to the state of low momentum $\hbar (k_1 - k_2)$, according to the diagram
\begin{center}
\vspace{\baselineskip}
\begin{tikzpicture}[thick]
\def\u{1}
\def\p{0.5*\u}
\node(t1) at (-\u,0) {};
\node (t2) at (-2*\u,0) {};
\node (t3) at (-3.5*\u,0) {};

\draw (0,0) node [above] {$0$} -- (t1);
\draw[snake it, <-] (t1)  -- + (\p,-\p) node[right] {$+k_1$};
\filldraw (t1) circle (0.05);
\draw[dashed] (t1) -- (t2);
\draw[snake it, ->] (t2) -- +(-\p,\p) node[right] {$+k_2$};
\filldraw (t2) circle (0.05);
\draw (t2) -- (t3) node [anchor = south] {$\hbar(k_1-k_2)$} ;
\end{tikzpicture}    
\end{center}

However, the probability amplitude 
\begin{align*}
&\bra{p= \hbar(k_1-k_2) } U_I(T,0) \ket{p=0}  = \\
&\quad   -i T\sum_n \frac{\Omega_{1,n}\Omega_{2,n} (\mathbf{R})}{4\delta_{1,n}} 
\left[
F( \frac{\omega_2 -\omega_1}2 T )- F(\frac{\delta_{2,n}}2 T)
\right]  
\end{align*}
vanishes as soon as $|\omega_2 - \omega_1|T\gg 1 $, again because of the energy mismatch $\simeq \hbar(\omega_2-\omega_1)$ between the final and initial states.

Instead, Bragg transitions towards low-momentum states $2\hbar(k_1 - k_2)$ are possible via four-photons processes
\begin{center}
\vspace{\baselineskip}
\begin{tikzpicture}[thick]
\def\u{1}
\def\p{0.5*\u}
\node(t1) at (-\u,0) {};
\node (t2) at (-2*\u,0) {};
\node (t3) at (-3*\u,0) {};
\node (t4) at (-4*\u,0) {};

\draw (0,0)  -- (t1);
\draw[snake it, <-] (t1)  -- + (\p,-\p) node[right] {$+k_1$};
\filldraw (t1) circle (0.05);
\draw[dashed] (t1) -- (t2);
\draw[snake it, ->] (t2) -- +(-\p,\p) node[right] {$-k_1$};
\filldraw (t2) circle (0.05);
\draw (t2) -- (t3); 
\filldraw (t3) circle (0.05);
\draw[snake it, <-] (t3)  -- + (\p,-\p) node[right] {$-k_2$};
\draw[dashed] (t3) -- (t4);
\filldraw (t4) circle (0.05);
\draw[snake it, ->] (t4) -- +(-\p,\p) node[right] {$+k_2$};
\draw (t4) -- +(-\u,0);
\end{tikzpicture}
\begin{tikzpicture}[thick]
\def\u{1}
\def\p{0.5*\u}
\node(t1) at (-\u,0) {};
\node (t2) at (-2*\u,0) {};
\node (t3) at (-3*\u,0) {};
\node (t4) at (-4*\u,0) {};

\draw (0,0)  -- (t1);
\draw[snake it, <-] (t1)  -- + (\p,-\p) node[right] {$-k_2$};
\filldraw (t1) circle (0.05);
\draw[dashed] (t1) -- (t2);
\draw[snake it, ->] (t2) -- +(-\p,\p) node[right] {$+k_2$};
\filldraw (t2) circle (0.05);
\draw (t2) -- (t3); 
\filldraw (t3) circle (0.05);
\draw[snake it, <-] (t3)  -- + (\p,-\p) node[right] {$+k_1$};
\draw[dashed] (t3) -- (t4);
\filldraw (t4) circle (0.05);
\draw[snake it, ->] (t4) -- +(-\p,\p) node[right] {$-k_1$};
\draw (t4) -- +(-\u,0);
\end{tikzpicture}
\end{center}
Notice that, since we consider subsequent absorption and emission of the photons of the same frequency, the probability amplitude of these two processes gives the same result that we obtained using perturbation theory from the optical potential [see Eq.\eqref{eq:amplitude_Q}]. By doing so, we neglect the following processes
\begin{center}
\begin{tikzpicture}[thick]
\def\u{1}
\def\p{0.5*\u}
\node(t1) at (-\u,0) {};
\node (t2) at (-2*\u,0) {};
\node (t3) at (-3*\u,0) {};
\node (t4) at (-4*\u,0) {};
\draw (0,0)  -- (t1);
\draw[snake it, <-] (t1)  -- + (\p,-\p) node[right] {$+k_1$};
\filldraw (t1) circle (0.05);
\draw[dashed] (t1) -- (t2);
\draw[snake it, ->] (t2) -- +(-\p,\p) node[right] {$+k_2$};
\filldraw (t2) circle (0.05);
\draw (t2) -- (t3); 
\filldraw (t3) circle (0.05);
\draw[snake it, <-] (t3)  -- + (\p,-\p) node[right] {$-k_2$};
\draw[dashed] (t3) -- (t4);
\filldraw (t4) circle (0.05);
\draw[snake it, ->] (t4) -- +(-\p,\p) node[right] {$-k_1$};
\draw (t4) -- +(-\u,0);
\end{tikzpicture}

\begin{tikzpicture}[thick]
\def\u{1}
\def\p{0.5*\u}
\node(t1) at (-\u,0) {};
\node (t2) at (-2*\u,0) {};
\node (t3) at (-3*\u,0) {};
\node (t4) at (-4*\u,0) {};
\draw (0,0)  -- (t1);
\draw[snake it, <-] (t1)  -- + (\p,-\p) node[right] {$+k_1$};
\filldraw (t1) circle (0.05);
\draw[dashed] (t1) -- (t2);
\draw[snake it, ->] (t2) -- +(-\p,\p) node[right] {$+k_2$};
\filldraw (t2) circle (0.05);
\draw (t2) -- (t3); 
\filldraw (t3) circle (0.05);
\draw[snake it, <-] (t3)  -- + (\p,-\p) node[right] {$+k_1$};
\draw[dashed] (t3) -- (t4);
\filldraw (t4) circle (0.05);
\draw[snake it, ->] (t4) -- +(-\p,\p) node[right] {$+k_2$};
\draw (t4) -- +(-\u,0);
\end{tikzpicture}
    
\begin{tikzpicture}[thick]
\def\u{1}
\def\p{0.5*\u}
\node(t1) at (-\u,0) {};
\node (t2) at (-2*\u,0) {};
\node (t3) at (-3*\u,0) {};
\node (t4) at (-4*\u,0) {};
\draw (0,0)  -- (t1);
\draw[snake it, <-] (t1)  -- + (\p,-\p) node[right] {$-k_2$};
\filldraw (t1) circle (0.05);
\draw[dashed] (t1) -- (t2);
\draw[snake it, ->] (t2) -- +(-\p,\p) node[right] {$-k_1$};
\filldraw (t2) circle (0.05);
\draw (t2) -- (t3); 
\filldraw (t3) circle (0.05);
\draw[snake it, <-] (t3)  -- + (\p,-\p) node[right] {$+k_1$};
\draw[dashed] (t3) -- (t4);
\filldraw (t4) circle (0.05);
\draw[snake it, ->] (t4) -- +(-\p,\p) node[right] {$+k_2$};
\draw (t4) -- +(-\u,0);
\end{tikzpicture}

\begin{tikzpicture}[thick]
\def\u{1}
\def\p{0.5*\u}
\node(t1) at (-\u,0) {};
\node (t2) at (-2*\u,0) {};
\node (t3) at (-3*\u,0) {};
\node (t4) at (-4*\u,0) {};
\draw (0,0)  -- (t1);
\draw[snake it, <-] (t1)  -- + (\p,-\p) node[right] {$-k_2$};
\filldraw (t1) circle (0.05);
\draw[dashed] (t1) -- (t2);
\draw[snake it, ->] (t2) -- +(-\p,\p) node[right] {$-k_1$};
\filldraw (t2) circle (0.05);
\draw (t2) -- (t3); 
\filldraw (t3) circle (0.05);
\draw[snake it, <-] (t3)  -- + (\p,-\p) node[right] {$-k_2$};
\draw[dashed] (t3) -- (t4);
\filldraw (t4) circle (0.05);
\draw[snake it, ->] (t4) -- +(-\p,\p) node[right] {$-k_1$};
\draw (t4) -- +(-\u,0);
\end{tikzpicture}
\end{center}
that give rise to amplitudes smaller by factors $\sim E_{2k_1}/(\hbar\omega_2-\hbar\omega_1)$ with respect to the two former diagrams.

\section{Band structure}

In this section, we provide some complementary information regarding the computation of the band structure of the BNSL.
We consider the BNSL potential defined in Sec.~\ref{sec:system}, with commensurate wavelengths obeying the condition $(n+1)\lambda_1 = n\lambda_2= \lambda_{-}$. For simplicity, here we consider the case $\phi_{+} = \phi_{-} = 0$. Then, by changing variables to $y=k_-x$, see Eq.~\eqref{eq:potential_2bis}, the Hamiltonian can be written as 
\begin{equation}
    \mathcal H_y = -\nabla_y^2 - \frac{V_1}{2} \cos(2ny) - \frac{V_2}{2} \cos\left[2(n+1)y\right],
    \label{eq:mathieu}
\end{equation}
where energies are measured in units of the recoil energy $E_{-}\equiv\hbar^{2}k_{-}^{2}/(2m)$ and we omitted a constant offset term.
Equation~\eqref{eq:mathieu} is a two-frequency Mathieu-Hill equation that has been studied extensively in the literature~\cite{trypogeorgosCotrappingDifferentSpecies2016,harteUltracoldAtomsMultiple2018,footTwofrequencyOperationPaul2018,jordan_nonlinear_2007,konenkov_matrix_2002,broersimo1998}.
According to Bloch's theorem, the wave functions can be expanded in a plane-wave basis $\phi_{q}(y) = e^{iqy/\hbar}u_{q}(y)$ where $q$ is the quasimomentum and the functions $u_{q}(y)$ have the same periodicity as the Hamiltonian, $u_{q}(y+\pi)=u_{q}(y)$.
Substituting this ansatz into the Hamiltonian and taking the Fourier transform of $u_{q}(y) = 1/\sqrt{2\pi}\sum_m c_{q,m}e^{i2my}$ we obtain the discrete matrix equation
\begin{align}
&(q - 2mk_{-})^2 c_{q,m} + \frac{V_1}{4}(c_{q,m+n} + c_{q,m-n}) 
\nonumber\\
&\quad + \frac{V_2}{4}(c_{q,m+n+1} + c_{q,m-n-1}) = \varepsilon_{qn}c_{q,m},
\label{eq:mathieu_q}
\end{align}
which can be diagonalized to compute the band structure.

\begin{figure}[t]
    \centering
    \includegraphics[]{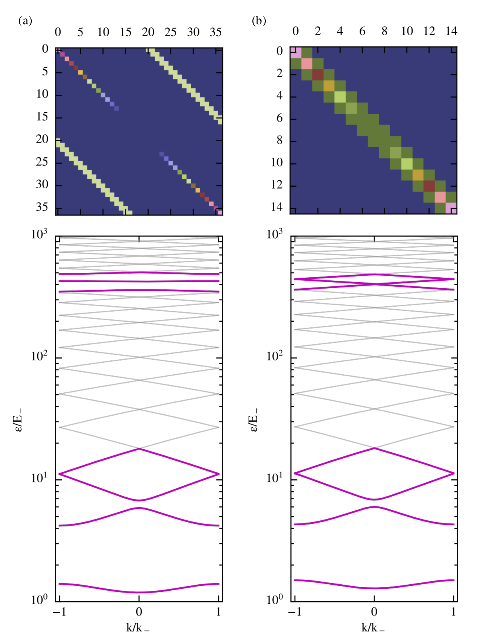}
    \caption{The Hamiltonian matrix and band structure corresponding to the periodic potential of (a) Eq.~\eqref{eq:mathieu_q} for a BNSL and (b) Eq.~\eqref{eq:mathieu_2} for a single lattice potential (see parameters in text).
    Although the lowest-energy bands are largely identical the underlying structure of the Hamiltonian matrices is completely different.
    Spectroscopically this difference appears only at the higher energy $n=20$ bands.
    }
    \label{fig:band_structure}
\end{figure}

Equation~\eqref{eq:mathieu_q} is a generalization of a single-lattice eigenvalue equation, and it reduces to it for $V_2=0$ and $n=1$. In particular, as we did in Fig.~\ref{fig3}, it is interesting to compare the general case with the single lattice effective potential of Eq.~\eqref{eq:veff}. In the dimensionless notation used above, the Hamiltonian of the latter reads 
\begin{equation}
    \mathcal H_y = -\nabla_y^2 - \frac{V_{1}^{2}}{16}\frac{E_{-}}{E_{B^{+}}}\cos(2y),
\label{eq:mathieu_2}
\end{equation}
which is obtained from Eq.~\eqref{eq:mathieu} by the substitution $V_{1}\to V_{1}^{2}(E_{-}/E_{B^{+}})/8$, along with setting $n=1$ and $V_{2}=0$.
The different structure of the Hamiltonian matrices for the two cases is shown in \aref{fig:band_structure}.
Both matrices have nonzero elements only on the diagonals, with the main diagonal being the same, as it corresponds to the kinetic energy term.
In the single-lattice case only the $\pm 1$ diagonals are populated, whereas in the case of the BNSL, both the $n$ and $n+1$ diagonals are populated.

Figure~\ref{fig:band_structure} shows the first 30 lattice bands obtained by diagonalizing \aref{eq:mathieu_q} in the reciprocal space, for the single-lattice in Eq.~\eqref{eq:mathieu_2} and BNSL in Eq.~\eqref{eq:mathieu_q}. These are obtained for (a) $n=20$ and $V_1=V_2=70$ (in units of $E_{-}$, as used here, which corresponds to $E_{B^{+}}/(n+1/2)^2$), and (b) for $n=1$, $V_1=1.457$, and $V_2=0$.
The energy gaps and bandwidth of the lowest energy bands are practically identical.
The origin of these low energy gaps, however, is different in the two cases:  in the single-lattice case, it arises from a direct two-photon process, where the photons have the same and opposite quasimomenta, while in the BNSL case, it corresponds to an effective Raman process at $k_2 - k_1$.
The direct processes for the BNSL can be seen as opening the gaps between the $n$ and $n+1$ bands.

\end{document}